\documentclass[twocolumn]{aastex631}

\shorttitle{Searches for XMP Galaxies}
\shortauthors{Miller et al.}

\graphicspath{{./}{figures/}}

\usepackage{natbib}
\usepackage{makecell}

\begin{document}

\title{Searches for Extremely Metal Poor Galaxies \\ using ALFALFA-selected Dwarf Galaxies\footnote{Based in part on observations obtained with the Hobby-Eberly Telescope, which is a joint project of the University of Texas at Austin, the Pennsylvania State University, Ludwig-Maximilians-Universität München, and Georg-August-Universität Göttingen.}}

\author[0000-0002-2901-5260]{John H. Miller Jr}
\affiliation{Department of Astronomy, Indiana University, 727 East Third Street, Bloomington, IN 47405, USA}
\affiliation{Minnesota Institute for Astrophysics, University of Minnesota, 116 Church Street, Minneapolis, MN 55455, USA}

\author[0000-0001-8483-603X]{John J. Salzer}
\affiliation{Department of Astronomy, Indiana University, 727 East Third Street, Bloomington, IN 47405, USA}

\author[0000-0001-9165-8905]{Steven Janowiecki}
\affiliation{McDonald Observatory, University of Texas, Austin, TX 78712, USA}

\author[0000-0001-5334-5166]{Martha P. Haynes}
\affiliation{Center for Astrophysics and Planetary Science, Space Science Building, Cornell University, Ithaca, NY 14853, USA}

\author[0000-0002-2954-8622]{Alec S. Hirschauer}
\affiliation{Space Telescope Science Institute, 3700 San Martin Dr., Baltimore, MD 21218, USA}


\begin{abstract}

We present a study of nearby dwarf galaxies selected from the ALFALFA blind \ion{H}{1} survey.  A primary goal of the project was to utilize a non-standard selection method with the hope of detecting previously unrecognized extremely metal-poor (XMP) galaxies.  The study was motivated by the recent discovery of two XMP galaxies -- Leo P and Leoncino -- which were both originally found via the ALFALFA survey.  We have obtained narrowband H$\alpha$ images for 42 dwarf systems, many of which are located in the local void in front of the Pisces-Perseus Supercluster.  Spectra for eleven of the best candidates resulted in the determination of metal abundances for ten of the systems.   None were found to be extremely metal poor, although one system (AGC 123350) was found to have an oxygen abundance of log(O/H)+12 = 7.46, or $\sim$6\% solar.  One of the galaxies in our sample exhibits a high oxygen abundance for its luminosity, suggesting the possibility that it may have a tidal origin.

\end{abstract}



\section{Introduction} \label{sec:intro}

Extremely metal-poor (XMP) galaxies are typically low-mass, gas-rich, star-forming systems that are observed in the local universe. The XMP categorization is given to galaxies with an oxygen abundance of log(O/H)+12 less than 7.35, or one twentieth solar metallicity \citep{mcquinn2020leoncino}. XMP galaxies have been utilized to probe star formation at low-metallicity \citep{annibali2022}, measure primordial abundances of the light elements to compare to Big Bang Nucleosynthesis models \citep{aver2021}, and set limitations on the initial mass function (IMF; \citealt{jev2018}). Studies of XMP galaxies aim to resolve the complexities of star formation in low-metallicity environments to help elucidate the formation of the first stars.

The discovery of XMP galaxies depends on observations of low-mass, low-luminosity galaxies in the local universe. Since metallicity is known to decrease with decreasing mass and luminosity \citep{Lequeux1979,skillman1989oxygen}, most efforts to discover XMPs have focused on low-mass systems.  A number of recent studies have searched for XMP candidates by selecting dwarf galaxies with extreme colors \citep[e.g.,][]{james2015, james2017, yang2017, hsyu2018}, but with limited success.

In the current study, we are utilizing \ion{H}{1} catalogs to investigate low \ion{H}{1} mass galaxies to potentially discover suitable XMP candidates.  This is motivated by recent discoveries of XMP galaxies found in \ion{H}{1}-selected low-mass samples (see below). The principal \ion{H}{1} catalog used to conduct these searches is the Arecibo Legacy Fast ALFA (ALFALFA) catalog.  The ALFALFA survey is a large scale, blind extragalactic \ion{H}{1} survey \citep{giovanelli2005arecibo,haynes2018arecibo}. The survey produced a comprehensive catalog of nearby, low-mass galaxies with \ion{H}{1} masses below 10$^8$ M$_{\odot}$, with masses as low as 10$^6$ M$_{\odot}$ \citep{giovanelli2005arecibo, giovanelli2005II}.  In total, the ALFALFA catalog contains about 31,500 extragalactic \ion{H}{1} sources.

Two projects that utilized the dwarf galaxy sub-sample of the ALFALFA catalog are the \textit{Survey for \ion{H}{1} in Extremely Low-mass Dwarfs} (SHIELD; \citealt{cannon2011survey,mcquinn2015}) and \textit{Ultra-Compact High Velocity Clouds} (UCHVC; \citealt{giovanelli2009,adams2013catalog}). During the course of these projects, optical observations indicated that two of these \ion{H}{1}-selected galaxies have very low metallicities.  Leo P \citep{rhode2013alfalfa,skillman2013alfalfa,mcquinn2015leop,aver2021} is a nearby (D = 1.62 Mpc) UCHVC system with log(O/H)+12 = 7.17 $\pm$ 0.04. It is currently the lowest-mass galaxy known that is actively forming stars, with M$_*$ = 5.6 $\times$ 10$^5$ M$_\odot$ \citep{mcquinn2015leop}.   AGC 198691 (a.k.a. the Leoncino Dwarf galaxy) is part of the SHIELD project, and has log(O/H)+12 = 7.06 $\pm$ 0.03 \citep{hirschauer2016alfalfa,mcquinn2020leoncino,aver2022}. Both galaxies are low-mass, gas-rich systems with ongoing star formation and are in relative isolation from any neighboring massive galaxies. 

These successes in discovering some of the most metal-poor objects known have demonstrated the efficacy of investigating the metallicity of low-mass ALFALFA-selected dwarf galaxies. The goal of the current study is to explore additional methods of discovering XMP galaxies using \ion{H}{1}-selected samples. In Section \ref{sec:2} we describe the sample selection and object list. Section \ref{sec:3} details the narrowband imaging observations and reductions, with the photometric results being discussed in Section \ref{sec:4}. The follow-up spectroscopic observations and metallicity determinations are described in Section \ref{sec:5}. Section \ref{sec:6} discusses the utility of our alternative method in discovering XMP galaxies and compares our results to known XMP galaxies. Section \ref{sec:7} summarizes our results.

\section{Sample Selection} \label{sec:2}

Our sample of galaxies was derived from the ALFALFA blind \ion{H}{1} survey \citep{giovanelli2005arecibo,haynes2018arecibo}. 
The ALFALFA survey represents the definitive catalog of low-mass gas-rich galaxies in the local Universe. Generally, galaxies with low \ion{H}{1} mass can be characterized as lower-mass dwarf galaxies. Previous studies have shown that some low-mass \ion{H}{1} detected sources are extremely metal-poor \citep{skillman2013alfalfa, hirschauer2016alfalfa}. In an effort to find additional XMP galaxies we are exploring the ALFALFA database to find additional candidates to be studied.

ALFALFA used a two-pass strategy to ensure that all \ion{H}{1} detections were correctly identified and were not caused by radio frequency interference or from other spurious signals. Most objects in the main ALFALFA catalog were observed with a signal-to-noise ratio of 6.5$\sigma$ or greater, although sources with lower detection significance but with a previously known redshift that matched the \ion{H}{1} detection were also included. Objects with a signal-to-noise ratio below the 6.5$\sigma$ threshold and without a matching prior redshift could not be reliably confirmed as \ion{H}{1} detections and were not included in the final published ALFALFA lists. 

The objects chosen for the current study were selected from among these lower signal-to-noise ratio objects found by ALFALFA.  They represent objects that were detected at the 4.0 - 6.5 $\sigma$ level in the original survey.  All of the objects we considered for our sample were re-observed with Arecibo after the completion of the ALFALFA survey using pointed observations to confirm that the putative ALFALFA detections were real \citep{o2019}.  The expectation for such a sample is that, at any given distance, the galaxies thus selected will have lower \ion{H}{1} masses than the objects in the main ALFALFA catalog.  This in turn means that they could have lower stellar masses and hence lower metal abundances.  Naturally, galaxies have a wide range of gas-mass fractions, hence this selection method does not necessarily translate into a sample with lower {\it stellar} masses.  Therefore, while this selection criterion has been implemented to increase the probability of discovering XMP galaxies, such a discovery is not a guarantee. 

Two additional selection criteria were imposed on our sample of ALFALFA galaxies.  First, the galaxies were chosen to lie within the Southern Galactic cap, in order to make them accessible during the Fall observing season.  All of the galaxies have right ascensions (RAs) between 22$^h$ and 3$^h$ and declinations (DEC) between 21$^\circ$ and 37$^\circ$.  Additionally, only galaxies with velocities less than 6000 km s$^{-1}$ are included in our sample.  This latter restriction serves two purposes for our project.  (1) The velocity limit reduces the effects of distance in \ion{H}{1} detections, effectively eliminating higher mass, higher metalicity galaxies. (2) The velocity limit also means that we are selecting galaxies that are mostly in the foreground of the Pisces-Perseus supercluster \citep{hg1986}, in the so-called Local Void.  Selecting a sample of galaxies located preferentially in low density environments may help in the detection of XMP galaxies, since at least some studies have suggested that low metallicity galaxies are preferentially found in voids \citep{pustilnik2016,kniasev2018,pustilnik2021}.  

The combination of a lower \ion{H}{1} detection threshold, low velocities, and a low-density environment suggests that the selected objects have the potential to include a number of small dwarf galaxies that are metal-poor. Our overall sample includes 78 ALFALFA galaxies that satisfy our selection criteria.

Table \ref{tab:table1} lists each galaxy in our ALFALFA-selected sample, ordered by increasing right ascension.  The table includes RA, DEC, observed velocity, velocity width of the \ion{H}{1} profile (and its error), \ion{H}{1} line flux integral, flow-model distance, and \ion{H}{1} mass.  Since many of the target galaxies are fairly nearby, we do not use a simple Hubble's Law computation for the distance.  Rather, we utilize the velocity flows model of \citet{masters2005galaxy} to estimate the distance.  Since the majority of our sample have velocities in excess of 3000 km s$^{-1}$, their flow-model distances should be close to their Hubble Law distances in most cases.

\startlongtable
\begin{deluxetable*}{crcrrccccc}
    \tabletypesize{\small}
    \phantom{$\pm$}
    \tablecaption{Initial ALFALFA Candidate List}
    \tablewidth{8cm}
    \tabletypesize{\footnotesize}
    \tablehead{\scshape{AGC \#} & \scshape{RA} & \scshape{DEC} & \scshape{Vel.} & \scshape{w50} & \scshape{\ion{H}{1} Flux} & \scshape{D} & \scshape{log(M$_{HI}$)} & \scshape{Priority}\\
    & [degrees] & [degrees] & [km s$^{-1}$] & [km s$^{-1}$] & [Jy km s$^{-1}$] & [Mpc] & [M$_{\odot}$] & \\
    (1) & (2) & (3) & (4) & (5) & (6) & (7) & (8) & (9)}
    \startdata
    104295 & 0.7748 & 34.6383 & 5167 & 26 $\pm$ 2 & 0.21 & 73.23 & 8.42 & 3 \\
    104299 & 1.2713 & 31.4619 & 4983 & 49 $\pm$ 5 & 0.29 & 70.52 & 8.53 & 4 \\
    104214 & 1.7658 & 33.3403 & 4694 & 41 $\pm$ 5 & 0.17 & 66.32 & 8.25 & 4 \\
    105229 & 2.0151 & 28.0508 & 4497 & 53 $\pm$ 2 & 0.17 & 63.45 & 8.21 & 3 \\
    105231 & 2.5003 & 29.5983 & 4681 & 51 $\pm$ 5 & 0.13 & 66.07 & 8.13 & 2 \\
    105234 & 3.0224 & 29.5178 & 4641 & 44 $\pm$ 7 & 0.15 & 65.42 & 8.18 & 3 \\
    105263 & 4.2655 & 29.0333 & 4805 & 54 $\pm$ 10 & 0.19 & 67.67 & 8.31 & 1 \\
    104827 & 5.2605 & 28.8867 & 4688 & 57 $\pm$ 3 & 0.49 & 65.90 & 8.70 & 2 \\
    104377 & 5.5059 & 27.4744 & 3900 & 56 $\pm$ 17 & 0.13 & 54.59 & 7.96 & 4 \\
    103425 & 10.2623 & 22.0419 & 812 & 41 $\pm$ 2 & 0.46 & 11.71 & 7.17 & 2 \\
    104446 & 10.5071 & 28.2619 & 4987 & 66 $\pm$ 5 & 0.20 & 69.78 & 8.36 & 5 \\
    104765 & 10.5118 & 25.3631 & 1654 & 36 $\pm$ 3 & 0.22 & 22.95 & 7.44 & 4 \\
    105273 & 11.2615 & 29.6675 & 5135 & 67 $\pm$ 3 & 0.22 & 71.88 & 8.43 & 2 \\
    104456 & 11.2671 & 30.9967 & 5045 & 36 $\pm$ 5 & 0.19 & 70.58 & 8.35 & 4 \\
    104785 & 11.2678 & 27.5989 & 5152 & 60 $\pm$ 9 & 0.29 & 72.11 & 8.55 & 3 \\
    102857 & 12.0101 & 31.5742 & 5075 & 52 $\pm$ 5 & 0.30 & 70.97 & 8.55 & 4 \\
    103503 & 12.0250 & 23.4225 & 2137 & 55 $\pm$ 3 & 0.56 & 29.38 & 8.06 & 2 \\
    102968 & 12.2616 & 28.3553 & 5007 & 57 $\pm$ 4 & 0.40 & 69.94 & 8.66 & 3 \\
    104494 & 13.5119 & 29.1331 & 4630 & 31 $\pm$ 2 & 0.30 & 64.36 & 8.45 & 4 \\
    101732 & 13.5166 & 31.9925 & 5365 & 40 $\pm$ 3 & 0.23 & 75.09 & 8.49 & 4 \\
    102973 & 13.7744 & 29.0042 & 4859 & 47 $\pm$ 8 & 0.40 & 67.67 & 8.64 & 5 \\
    102859 & 14.7576 & 30.5922 & 4687 & 67 $\pm$ 5 & 0.40 & 65.12 & 8.60 & 3 \\
    115349 & 15.0047 & 30.6108 & 5154 & 37 $\pm$ 3 & 0.34 & 71.90 & 8.62 & 2 \\
    115382 & 17.5161 & 31.7317 & 4830 & 48 $\pm$ 12 & 0.19 & 67.01 & 8.30 & 3 \\
    116208 & 17.7552 & 34.6992 & 4934 & 47 $\pm$ 4 & 0.25 & 68.54 & 8.44 & 3 \\
    115447 & 21.5159 & 31.7431 & 4663 & 68 $\pm$ 9 & 0.16 & 64.33 & 8.19 & 3 \\
    116230 & 22.7522 & 23.8647 & 3421 & 70 $\pm$ 12 & 0.23 & 46.51 & 8.07 & 3 \\
    114082 & 25.0246 & 30.0775 & 4279 & 58 $\pm$ 4 & 0.33 & 58.57 & 8.43 & 3 \\
    113863 & 25.7658 & 26.0042 & 3898 & 67 $\pm$ 4 & 0.33 & 53.09 & 8.34 & 1 \\
    115320 & 27.0210 & 22.1136 & 2955 & 60 $\pm$ 5 & 0.37 & 39.82 & 8.14 & 3 \\
    113305 & 28.2713 & 26.2631 & 2813 & 66 $\pm$ 5 & 0.54 & 37.78 & 8.26 & 1 \\
    123069 & 30.5144 & 31.2656 & 4725 & 63 $\pm$ 6 & 0.21 & 64.81 & 8.32 & 4 \\
    124660 & 31.0050 & 27.3911 & 5183 & 48 $\pm$ 6 & 0.20 & 71.33 & 8.46 & 2 \\
    124021 & 31.0108 & 22.1694 & 4887 & 57 $\pm$ 3 & 0.42 & 67.23 & 8.65 & 3 \\
    125165 & 31.2633 & 31.8422 & 4848 & 61 $\pm$ 4 & 0.26 & 66.58 & 8.43 & 2 \\
    123182 & 32.2670 & 28.9261 & 4983 & 33 $\pm$ 2 & 0.30 & 68.54 & 8.52 & 4 \\
    124697 & 33.0199 & 31.3972 & 5122 & 69 $\pm$ 7 & 0.29 & 70.55 & 8.53 & 2 \\
    125180 & 33.2587 & 31.7042 & 5002 & 54 $\pm$ 6 & 0.28 & 68.78 & 8.49 & 2 \\
    123239 & 34.0211 & 29.3256 & 5249 & 49 $\pm$ 5 & 0.29 & 72.40 & 8.55 & 4 \\
    125209 & 34.2572 & 29.4403 & 5242 & 64 $\pm$ 2 & 0.93 & 72.30 & 9.06 & 1 \\
    123191 & 34.7610 & 29.7086 & 1798 & 25 $\pm$ 2 & 0.17 & 23.95 & 7.36 & 4 \\
    122933 & 34.7718 & 24.2969 & 5164 & 68 $\pm$ 5 & 0.45 & 71.17 & 8.73 & 3 \\
    123087 & 35.2649 & 31.5736 & 5075 & 44 $\pm$ 6 & 0.22 & 69.81 & 8.40 & 5 \\
    124722 & 35.5200 & 31.7692 & 5098 & 59 $\pm$ 7 & 0.32 & 70.14 & 8.57 & 5 \\
    123089 & 35.7697 & 30.3125 & 5154 & 68 $\pm$ 6 & 0.35 & 70.96 & 8.62 & 4 \\
    123350 & 36.7544 & 26.6389 & 898 & 35 $\pm$ 2 & 0.35 & 12.25 & 7.09 & 2 \\
    122939 & 36.7555 & 23.9608 & 1524 & 56 $\pm$ 3 & 0.44 & 20.31 & 7.63 & 3 \\
    124040 & 37.0117 & 22.2319 & 4342 & 51 $\pm$ 4 & 0.18 & 59.20 & 8.17 & 2 \\
    123200 & 38.7590 & 29.4469 & 1677 & 51 $\pm$ 3 & 0.50 & 22.31 & 7.77 & 5 \\
    124755 & 40.0061 & 29.9761 & 5209 & 56 $\pm$ 7 & 0.18 & 71.71 & 8.34 & 4 \\
    125544 & 40.0131 & 29.1647 & 5337 & 39 $\pm$ 4 & 0.14 & 73.61 & 8.25 & 4 \\
    123099 & 41.2639 & 31.3556 & 5180 & 42 $\pm$ 2 & 0.38 & 71.27 & 8.66 & 5 \\
    125524 & 41.2675 & 31.4097 & 5180 & 60 $\pm$ 6 & 0.16 & 71.26 & 8.28 & 5 \\
    125245 & 42.7665 & 33.7800 & 4346 & 70 $\pm$ 2 & 0.72 & 59.02 & 8.77 & 5 \\
    123101 & 44.5226 & 30.5675 & 3260 & 64 $\pm$ 3 & 0.44 & 43.62 & 8.30 & 3 \\
    124773 & 44.7550 & 34.9914 & 4896 & 48 $\pm$ 3 & 0.12 & 67.04 & 8.10 & 5 \\
    123213 & 44.7564 & 28.5492 & 5637 & 58 $\pm$ 3 & 0.42 & 78.04 & 8.78 & 3 \\
    123254 & 44.7573 & 24.8119 & 4652 & 56 $\pm$ 3 & 0.41 & 63.62 & 8.59 & 3 \\
    131439 & 45.0161 & 25.7836 & 303 & 19 $\pm$ 2 & 0.14 & 4.61 & 5.74 & 3 \\
    323400 & 332.0007 & 29.2575 & 2884 & 56 $\pm$ 17 & 0.14 & 43.88 & 7.80 & 2 \\
    322496 & 332.0069 & 30.6667 & 1115 & 45 $\pm$ 1 & 0.50 & 18.38 & 7.60 & 3 \\
    323421 & 336.2537 & 36.4606 & 1287 & 46 $\pm$ 2 & 0.79 & 20.55 & 7.90 & 3 \\
    322064 & 336.5007 & 22.9056 & 1275 & 27 $\pm$ 2 & 0.28 & 20.24 & 7.43 & 2 \\
    322522 & 337.7641 & 31.0839 & 3361 & 68 $\pm$ 7 & 0.34 & 49.97 & 8.30 & 2 \\
    322635 & 344.7545 & 34.4450 & 875 & 32 $\pm$ 3 & 0.26 & 14.14 & 7.09 & 2 \\
    333446 & 346.5010 & 31.1208 & 5916 & 46 $\pm$ 4 & 0.37 & 85.19 & 8.80 & 1 \\
    336092 & 349.0089 & 34.3089 & 1932 & 27 $\pm$ 2 & 0.25 & 28.53 & 7.68 & 4 \\
    333330 & 349.5042 & 25.8336 & 5781 & 34 $\pm$ 4 & 0.18 & 82.96 & 8.47 & 2 \\
    334540 & 349.7623 & 22.5244 & 5037 & 62 $\pm$ 3 & 0.66 & 72.35 & 8.91 & 3 \\
    332921 & 350.2670 & 27.8578 & 3664 & 66 $\pm$ 5 & 0.42 & 52.76 & 8.45 & 3 \\
    336095 & 350.2735 & 26.4983 & 5999 & 50 $\pm$ 3 & 0.48 & 85.99 & 8.92 & 1 \\
    333336 & 353.5093 & 25.5419 & 3719 & 69 $\pm$ 6 & 0.43 & 53.26 & 8.46 & 5 \\
    333337 & 354.5222 & 24.2314 & 720 & 21 $\pm$ 3 & 0.24 & 11.25 & 6.86 & 2 \\
    336100 & 354.7515 & 31.4736 & 3923 & 52 $\pm$ 8 & 0.30 & 56.22 & 8.23 & 3 \\
    333466 & 354.7589 & 31.7014 & 4153 & 42 $\pm$ 5 & 0.21 & 59.29 & 8.24 & 5 \\
    335430 & 355.5151 & 31.1764 & 2302 & 70 $\pm$ 3 & 0.12 & 33.07 & 7.49 & 4 \\
    334580 & 356.0229 & 23.6639 & 5430 & 48 $\pm$ 2 & 0.55 & 77.35 & 8.89 & 4 \\
    334595 & 359.2549 & 22.3592 & 4353 & 53 $\pm$ 6 & 0.33 & 61.66 & 8.47 & 3 \\
    \enddata 
    \tablecomments{The initial ALFALFA-selected object list. Due to the proximity of the galaxies, flow-model distances (column 8) were obtained from \cite{masters2005galaxy} and have an error of $\pm$ 2.33 Mpc.}
    \label{tab:table1}
\end{deluxetable*}

\begin{figure*}
    \centering
    \epsscale{1.10}
    \plotone{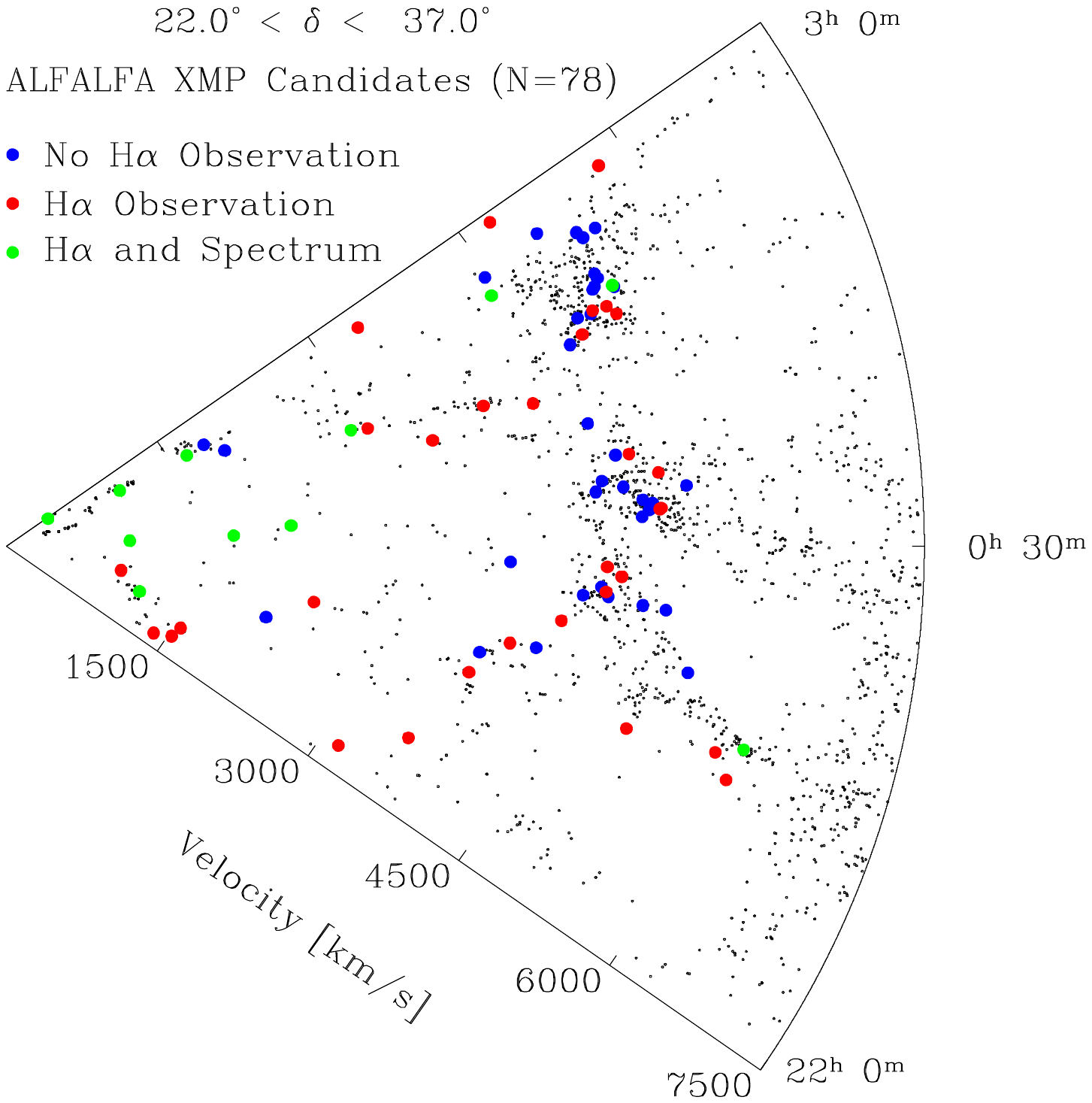}
    \caption{Cone diagram showing the spatial distribution of galaxies in our survey region.  The small black dots represent galaxies from various redshift surveys, including the ALFALFA survey \citep{haynes2018arecibo}, collected from the literature and included in the Arecibo Galaxy Catalog (AGC) which is maintained by MPH.  These galaxies are used to define the large-scale structure in this region of the nearby universe.  The current sample of XMP candidates selected from among the ALFALFA survey low SNR subsample are shown as colored circles.  Blue circles represent candidates that were not observed as part of our narrowband imaging project, red circles were galaxies for which H$\alpha$ NB images were obtained, while green circles were observed in H$\alpha$ as well as spectroscopically.}
    \label{fig:cone}
\end{figure*}

In Figure~\ref{fig:cone} we present a cone diagram that reveals the spatial distribution of galaxies located in the region of space from which the XMP candidates were selected.   The RA range of our sample is between 22$^h$ and 3$^h$, while the declination range is between 22$^\circ$ and 37$^\circ$.  The small dots in the diagram represent galaxies whose redshifts have been measured in previous studies; many of these were first measured in the ALFALFA survey \citep[e.g.,][]{haynes2018arecibo}.  These objects are used to define the overall distribution of galaxies in our field.  We note the high density ridge of galaxies between approximately 5000 and 6000 km s$^{-1}$, which is part of the Pisces-Perseus Supercluster \citep[PPS;][]{hg1986}, as well as the large lower density volume in the foreground of the PPS.

The ALFALFA-selected XMP candidates sample are shown in Figure~\ref{fig:cone} as colored circles.  The blue circles are those objects that were not observed as part of our narrowband imaging project (see \S~\ref{sec:3}), while the red and green circles denote galaxies for which H$\alpha$ images have been obtained.  The green circles indicate objects for which we were able to acquire follow-up spectra, as described in \S~\ref{sec:5}.  Many of the XMP candidates with velocities lower than 4500 km s$^{-1}$ are located in very low-density regions.  Our spectroscopic observations tended to include galaxies at these lower velocities; eight of the eleven objects for which spectra were obtained have velocities of 3000 km s$^{-1}$ or less.

The ALFALFA-selected objects were visually inspected using the Sloan Digital Sky Survey (SDSS; \citealt{york2000sloan}) using Data Release 15 (DR15; \citealt{aguado2019fifteenth}) to determine each galaxy's observational priority. Previous detections of extremely metal poor galaxies \citep[e.g.,][]{skillman2013alfalfa,hirschauer2016alfalfa} demonstrate that the most metal-poor objects tend to be compact systems with the presence of optical blue knots due to recent star formation.  The goal of the visual prioritization was to determine which of the target galaxies show evidence for the presence of strong current star formation that would result in a solid detection in our H$\alpha$ imaging observations. 

\begin{figure}
    \centering
    \epsscale{1.12}
    \plotone{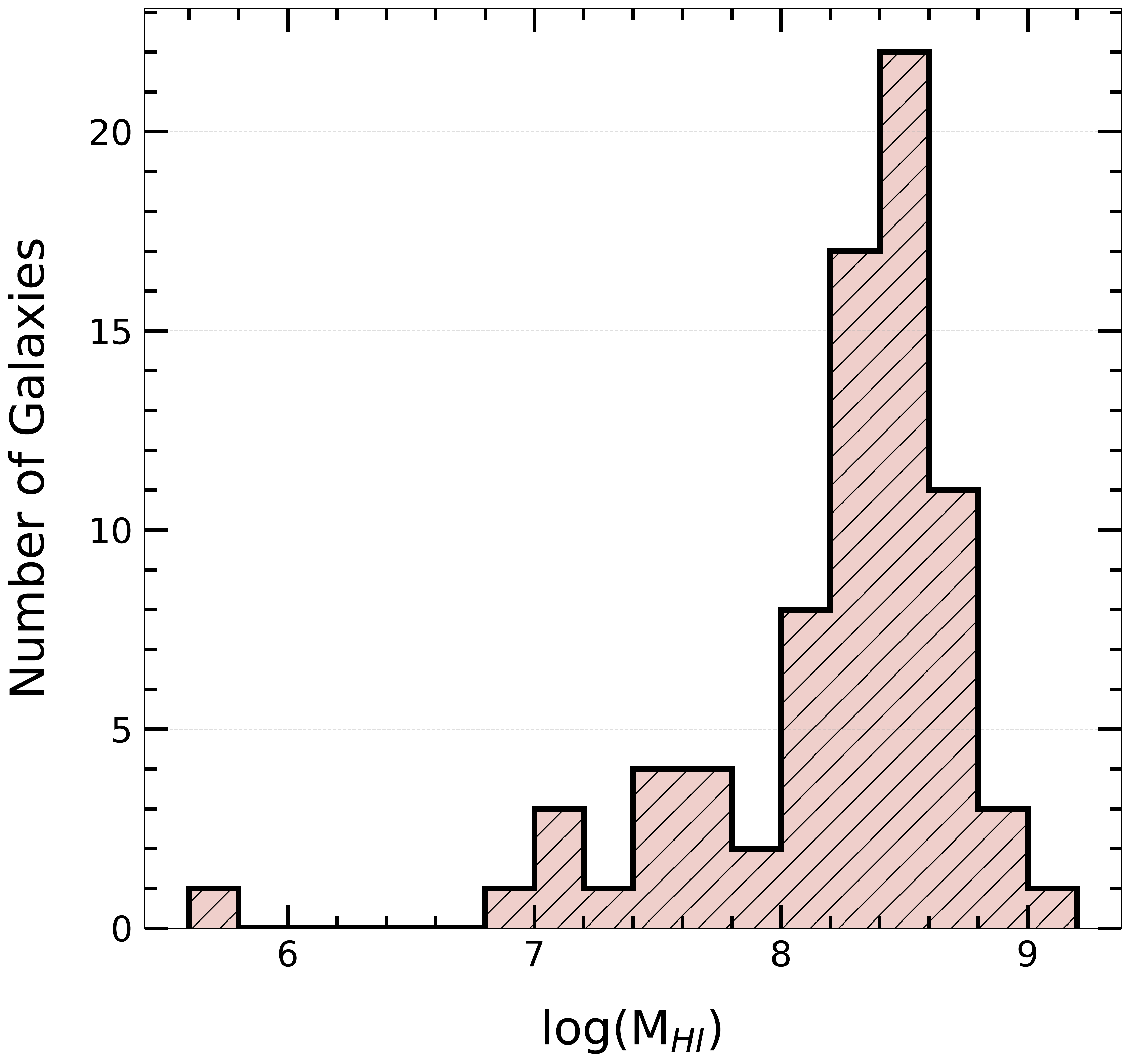}
    \caption{A histogram of the log of M$_{HI}$ for the ALFALFA selected galaxies. The median and average values of log(M$_{HI}$) are 8.36 and 8.24, respectively.}
    \label{fig:MHI}
\end{figure}

Based on our visual inspection, the galaxies in our sample were ranked from one to five, with one being the highest priority. Priority one objects contain compact, blue knots indicating that the galaxies likely have active star formation and a strong nebular spectrum suitable for measuring their metal abundance. Priority two and three objects represent galaxies with lower levels of star formation, as indicated by less extreme (less blue) optical colors and less prominent star-formation knots. Finally, objects in priority four and five contain redder galaxies that have little to no indication of active star formation. Although the attributes for each priority class are reasonably well defined, an object's priority does not necessarily correlate one-to-one with strength of H$\alpha$ emission due to the idiosyncratic nature of galaxies and difficulty of visually analyzing optical star-formation indicators.  Each galaxy in the sample was visually inspected independently by two of us (JHM and JJS), and a priority class was determined by averaging the two lists.  The final priority classes are listed for each galaxy in the final column of Table~\ref{tab:table1}, and are summarized in Table~\ref{tab:observed}.

Figure \ref{fig:MHI} shows the distribution of log(M$_{HI}$) for our initial target list selected from ALFALFA using the criteria described above.  The mean and median values of log(M$_{HI}$) are 8.24 M$_\odot$ and 8.36 M$_\odot$, respectively.   While all of the galaxies in our sample have \ion{H}{1} masses below 10$^9$ M$_\odot$, a large majority (62 of 78) have masses between 10$^8$ and 10$^9$ M$_\odot$.  These M$_{HI}$ masses are fairly normal for dwarf galaxies, and, for our current purposes, seem less likely to produce XMP galaxy candidates.  The 16 ALFALFA targets with \ion{H}{1} masses below 10$^8$ M$_\odot$ appear to be more interesting in that regard.

\section{Imaging Observations and Analysis} \label{sec:3}

\subsection{Observations}

Narrowband H$\alpha$ imaging observations were obtained using the Half-Degree Imager (HDI) on the WIYN 0.9m telescope\footnote{The 0.9m telescope is operated by WIYN Inc. on behalf of a consortium of partner universities and organizations. WIYN Inc. is a joint partnership of the University of Wisconsin at Madison, Indiana University, and the NOIRLab.}, located at the Kitt Peak National Observatory (KPNO) near Tucson, Arizona. Observations were carried out on six consecutive nights in 2019 from October 22nd through October 27th. Data were obtained on five of the six nights; no data were obtained on October 27th due to unsatisfactory weather conditions. The sky conditions for all five nights where data were obtained were generally photometric. 

The WIYN 0.9m telescope has a set of five narrowband H$\alpha$ filters. The velocity ranges and filter band-pass information is tabulated in \cite{van2016alfalfa}. The proper narrowband filter to use for each galaxy was determined from its observed heliocentric radial velocity obtained from ALFALFA. The galaxies selected for observation have a lower heliocentric velocity limit of 303 km s$^{-1}$ and an upper limit of 5999 km s$^{-1}$. Therefore, four narrowband H$\alpha$ filters were utilized to cover the full velocity range. Galaxies with similar redshift were observed on the same night to maximize our observing efficiency.

\begin{table}[t]
    \caption{Summary of Priority Classes}
    \movetableright=12pt
    \begin{tabular}{ccc} \hline \hline
    Priority & \# Galaxies & \# Observed \\ \hline
    1 & 6 & 6 \\
    2 & 18 & 18\\
    3 & 25 & 17\\
    4 & 18 & 2\\
    5 & 11 & 0\\  \hline
    \end{tabular}
    \label{tab:observed}
\end{table}

Three images were obtained for each target galaxy: one broadband R image bounded by two narrowband H$\alpha$ images. The images were observed in the standard H$\alpha$-R-H$\alpha$ sequence with a 45 minute total exposure time: 20 minutes per H$\alpha$ image and 5 minutes per R image. The HDI CCD contains 4096 $\times$ 4112 pixels with a pixel size of 15 $\mu$m. At the cassegrain focus of the WIYN 0.9m, the field of view spans 29$\arcmin$ $\times$ 29$\arcmin$\ with an image scale of 0\farcs 43 pixel$^{-1}$.  

Each night, approximately 20 bias images and five dome flats, per filter, were imaged to be used in standard image processing. Furthermore, images of spectrophotometric standard stars were taken each night to calibrate the flux measurements. 

The allotted observation time on the 0.9m allowed us to observe 43 of the 78 ALFALFA candidate galaxies. The observation list was determined by object priority and organized nightly by narrowband filter. Table \ref{tab:observed} summarizes the degree of success of our imaging observations.  Listed are the number of ALFALFA target galaxies as a function of priority class, as well as the number included in our imaging observations.  We were able to observe nearly all of the targets in the top three priority classes (41 of 49, or 84\%), including all 24 in the top two groups.  Only 2 of the 29 galaxies in the two lower priority groups were imaged (7\%).  These latter two groups are expected to include galaxies with weak or nonexistent nebular emission.  As seen in Figure~\ref{fig:cone}, many of these unobserved, low priority targets are located near the upper end of the redshift distribution of our sample, and are located within the higher density PPS.

\subsection{Image Processing and Analysis}

Image processing followed standard procedures. The broadband and narrowband images for each of the galaxies and standard stars had overscan and mean bias image corrections applied. They were then corrected for pixel-to-pixel sensitivity variations using mean dome flat field images. Additionally, spurious high energy particle events were removed using the \texttt{L.A. COSMIC} program for all images except the standards \citep{van2001cosmic}. 

The images were then put through a series of \texttt{IRAF} scripts (referred to as the `pipeline’) that was developed by Arthur Sugden to analyze ALFALFA images \citep{van2016alfalfa}.  The pipeline has four main steps that it uses to create a continuum subtracted H$\alpha$ image: image alignment, full width at half maximum (FWHM) equalization, flux scaling, and continuum subtraction. The first step aligns the two narrowband images with the broadband image using a number of well exposed stars. Then, the program equalizes the image resolutions by smoothing the images with smaller FWHMs to match the image with the largest FWHM. The penultimate step in the program determines the average flux of a series of stars in all three images and scales the R-band image and the second narrowband image to match the average flux of the first narrowband image. Finally, the code subtracts the scaled R-band image from the two H$\alpha$ narrowband images and combines the two images to create a final continuum-subtracted H$\alpha$ image.

The pipeline reduced images were then measured using a program to determine the broadband and narrowband flux of the target galaxies. The program plots a series of apertures centered at a user-designated location near the center of emission. Sources other than the target galaxy within or near the selected aperture can be masked by the user.  Based on the curve-of-growth generated from the multi-aperture photometry, the user selects the aperture that maximizes the enclosed flux while minimizing the formal uncertainty.

Our measurement process yields instrumental magnitudes for both the R-band and narrowband continuum-subtracted images for each galaxy observed. Two photometry codes, one each for the broadband and narrowband measurements, were used to take these instrumental magnitudes and convert them into calibrated apparent magnitudes by applying the relevant atmospheric extinction corrections and zero-point constants derived from the standard star observations. The narrowband photometry code additionally accounts for the wavelength-dependent transmission of the narrowband filters, using the known redshift of the galaxies.

\section{Results of Narrow-band Imaging} \label{sec:4}

\begin{figure}[t]
    \centering
    \epsscale{1.15}
    \plotone{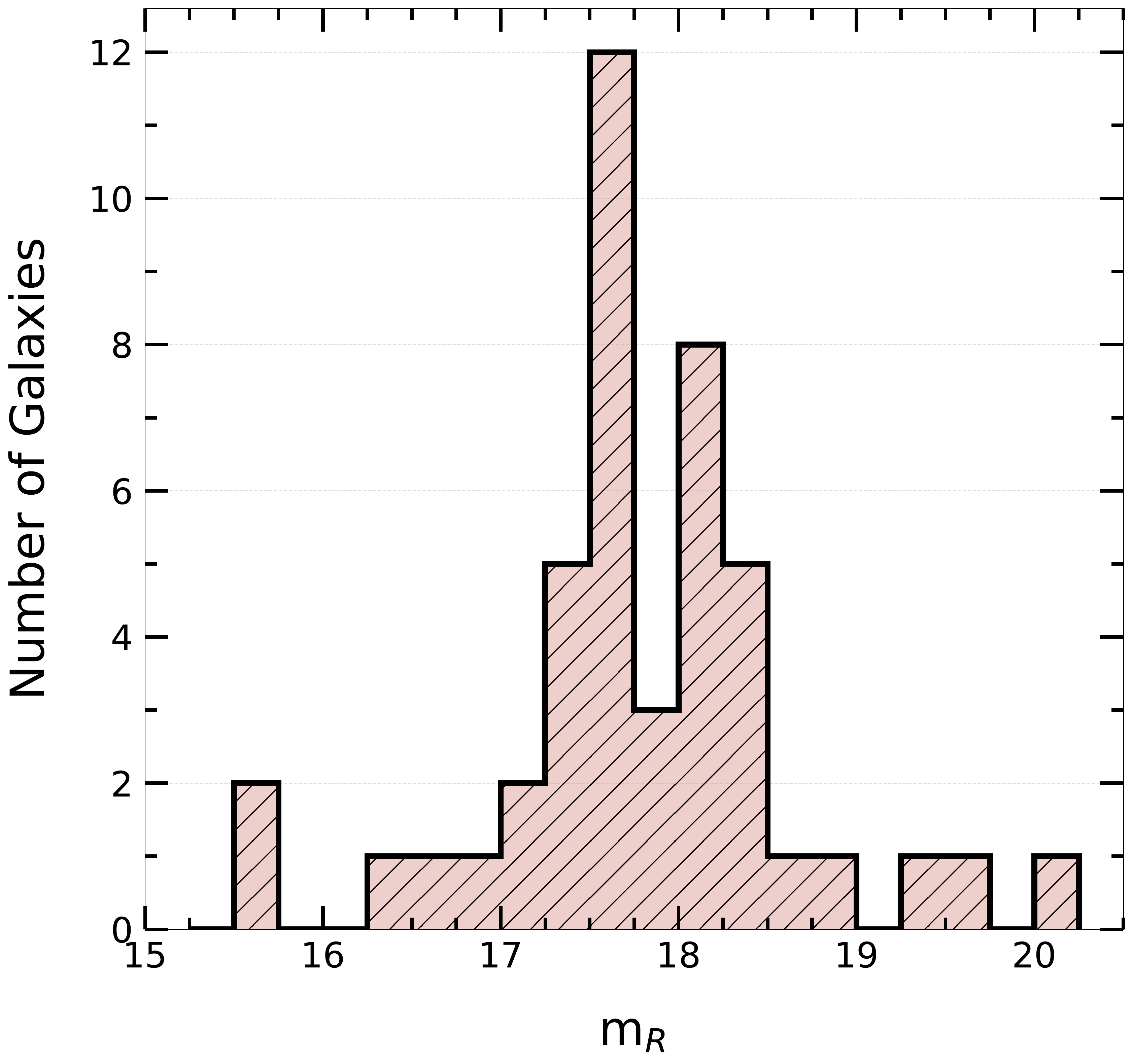}
    \caption{A histogram of the apparent magnitudes for the observed galaxies. The median and average values of m$_R$ are 17.71 and 17.79, respectively.}
    \label{fig:mR}
\end{figure}
    \begin{figure}[t]
    \centering
    \epsscale{1.15}
    \plotone{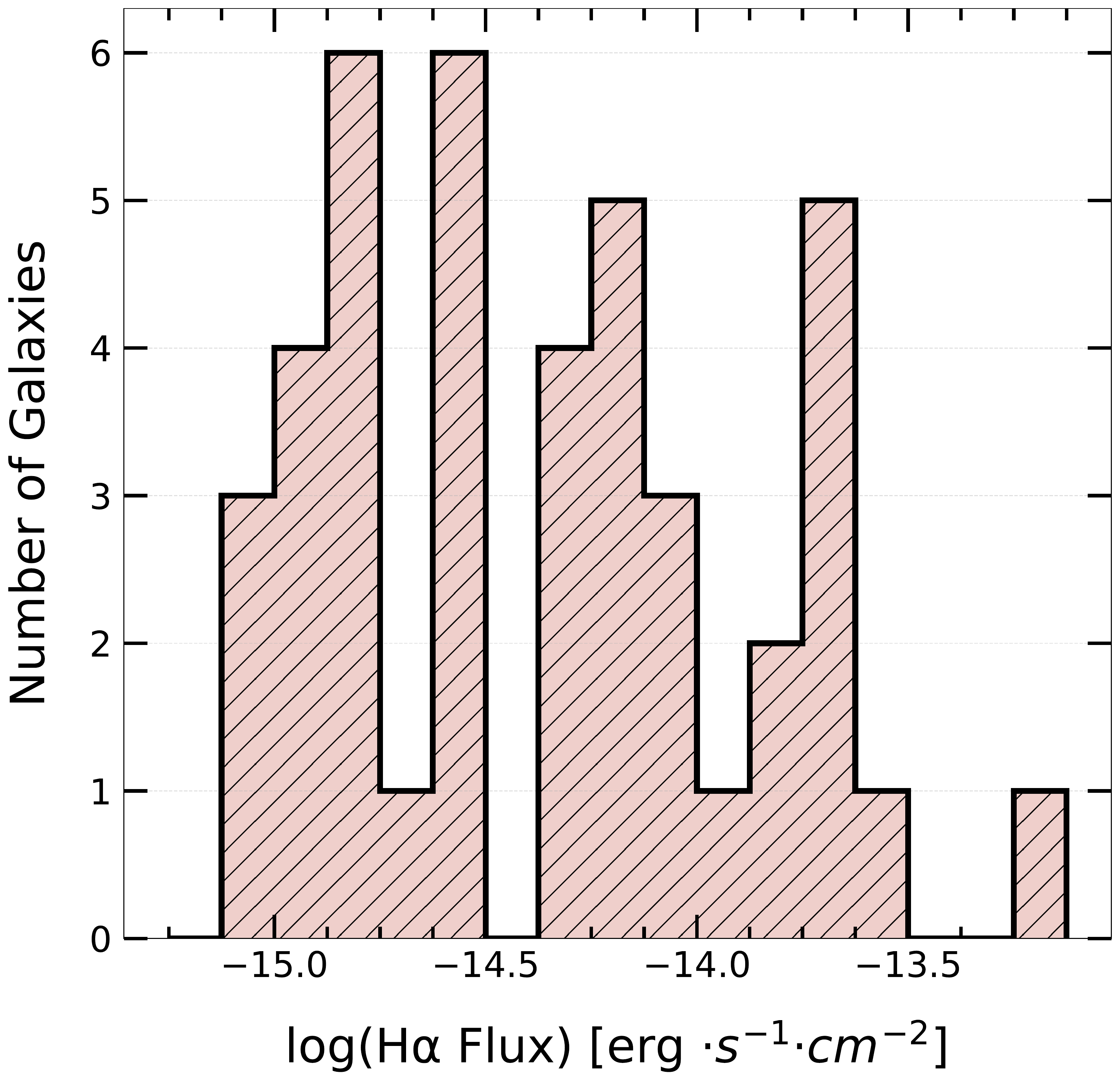}
    \caption{A histogram of the log of the H$\alpha$ flux for the observed galaxies. The median and average values of the H$\alpha$ flux for the XMP galaxies are 0.285 and 0.778 x10$^{-14}$ ergs s$^{-1}$ cm$^{-2}$, respectively. Three observed galaxies, AGC 322064, AGC 123101, and AGC 334595, were not included due to non-detections in H$\alpha$.}
    \label{fig:logFHA}
\end{figure}

A total of 43 dwarf galaxies were observed with the WIYN 0.9m telescope in R and H$\alpha$. Table \ref{tab:table2} presents the measured and calculated physical parameters from our photometric observations. The first column lists each observed galaxy's AGC number in RA order, while the second column lists the narrowband filters used for each galaxy using the filter designations from \cite{van2016alfalfa}.

Column 4 lists the apparent R-band magnitudes and their uncertainties.  These were determined from the R-band flux enclosed within the selected aperture diameter (Column 3) centered on the objects. Figure \ref{fig:mR} presents the distribution of these apparent R-band magnitudes for our sample. The observed galaxies have apparent magnitudes from 20.22 to 15.55, with a median and average values of 17.71 and 17.79, respectively. Approximately 80\% of the sample's apparent magnitudes lie between 17 and 18.5 magnitudes, creating a heavily peaked brightness distribution.

Column 6 lists the H$\alpha$ flux and error (in units of 10$^{-14}$ ergs s$^{-1}$ cm$^{-2}$), respectively, that were determined from the H$\alpha$ flux enclosed within the selected aperture diameter (Column 5) centered on the objects. Figure \ref{fig:logFHA} shows the distribution of H$\alpha$ flux values for the observed galaxies. Notably, the galaxy with the largest H$\alpha$ flux is AGC 113305 with (6.42 $\pm$ 0.13) $\times$ 10$^{-14}$ ergs s$^{-1}$ cm$^{-2}$. For reference, the ALFALFA detected XMP galaxy Leo P was found to have an H$\alpha$ flux of (1.71 $\pm$ 0.03) $\times$ 10$^{-14}$ ergs s$^{-1}$ cm$^{-2}$ \citep{rhode2013alfalfa}. We note that for three sources the signal-to-noise in the H$\alpha$ detection is extremely low and consider these to be non-detections, treating their fluxes as zero in the remainder of the paper.

The absolute R-band magnitude (Column 7) for each galaxy was calculated with the standard distance modulus equation corrected for Galactic absorption. The Galactic R-band absorption values were obtained from the NASA/ IPAC Extragalactic Database (NED)\footnote{http://ned.ipac.caltech.edu/} and the flow-model distances were obtained from \cite{masters2005galaxy} (Table \ref{tab:table1}). The derived H$\alpha$ luminosities (L(H$\alpha$); Column 8) for each galaxy have been corrected for Milky Way absorption using the same Galactic R-band absorption values used for the absolute R-band magnitude calculations. Finally, the star formation rate (SFR; Column 9) for each galaxy was calculated with the convention established in \cite{kennicutt1998star}, where SFR = 7.9 $\cdot$ 10$^{-42}$ $\cdot$ L(H$\alpha$) M$_{\odot}$ yr$^{-1}$.  We note that L(H$\alpha$) and SFR values have not been corrected for absorption internal to the target galaxy because we lack a spectroscopic absorption measurement in most cases. Since our targets are all dwarf galaxies, we expect the level of internal absorption to be modest.

\begin{deluxetable*}{lcccrcccc}
    \tabletypesize{\small}
    \tablehead{}
    \caption{Measured and Derived Physical Parameters}
    \tablewidth{8cm}
    \tablehead{\scshape{AGC$\#$} & \scshape{Filter} & \scshape{D$_R$} & \scshape{R} & \scshape{D$_{H\alpha}$} &  \scshape{F(H$\alpha$)} & \scshape{M$_R$} & \scshape{L(H$\alpha$)} & \scshape{log(SFR)} \\
    & & [arcsec] & [mag] & [arcsec] & [x10$^{-14}$ erg s$^{-1}$cm$^{-2}$] & [mag] & [erg s$^{-1}$] & [M$_{\odot}$yr$^{-1}$] \\
    (1) & (2) & (3) & (4) & (5) & (6) & (7) & (8) & (9)
    }
    \startdata
    105231  & 3 & 20 & 18.66 $\pm$ 0.02 & 12 & 0.12 $\pm$ 0.02 & -15.53 & 7.06 x 10$^{38}$ & -2.25 \\
    105263  & 3 & 30 & 17.57 $\pm$ 0.02 & 16 & 0.79 $\pm$ 0.03 & -16.66 & 4.63 x 10$^{39}$ & -1.44 \\
    104827B & 3 & 14 & 20.22 $\pm$ 0.20 & 6  & 0.25 $\pm$ 0.02 & -13.95 & 1.40 x 10$^{39}$ & -1.96 \\
    103425  & 1 & 36 & 15.55 $\pm$ 0.02 & 12 & 2.12 $\pm$ 0.05 & -14.91 & 3.87 x 10$^{38}$ & -2.52 \\
    104765  & 2 & 18 & 18.28 $\pm$ 0.05 & 8  & 0.25 $\pm$ 0.02 & -13.60 & 1.65 x 10$^{38}$ & -2.88 \\
    105273  & 3 & 24 & 17.57 $\pm$ 0.04 & 12 & 0.48 $\pm$ 0.03 & -16.82 & 3.29 x 10$^{39}$ & -1.59 \\
    104785  & 3 & 22 & 17.66 $\pm$ 0.02 & 12 & 0.17 $\pm$ 0.02 & -16.76 & 1.22 x 10$^{39}$ & -2.02 \\
    103503  & 2 & 34 & 15.65 $\pm$ 0.03 & 20 & 2.23 $\pm$ 0.07 & -16.78 & 2.50 x 10$^{39}$ & -1.70 \\
    115349  & 3 & 24 & 18.44 $\pm$ 0.07 & 10 & 0.19 $\pm$ 0.02 & -15.97 & 1.31 x 10$^{39}$ & -1.99 \\
    116208  & 3 & 20 & 18.39 $\pm$ 0.03 & 8  & 0.15 $\pm$ 0.02 & -15.93 & 9.34 x 10$^{38}$ & -2.13 \\
    116230  & 2 & 20 & 17.35 $\pm$ 0.03 & 12 & 0.26 $\pm$ 0.01 & -16.19 & 8.11 x 10$^{38}$ & -2.19 \\
    114082  & 3 & 20 & 17.98 $\pm$ 0.04 & 8  & 0.18 $\pm$ 0.03 & -15.96 & 7.95 x 10$^{38}$ & -2.20 \\
    113863  & 3 & 22 & 17.47 $\pm$ 0.02 & 16 & 0.66 $\pm$ 0.03 & -16.38 & 2.72 x 10$^{39}$ & -1.67 \\
    115320  & 2 & 40 & 16.94 $\pm$ 0.05 & 8  & 0.29 $\pm$ 0.02 & -16.25 & 6.44 x 10$^{38}$ & -2.29 \\
    113305  & 2 & 22 & 17.28 $\pm$ 0.03 & 14 & 6.42 $\pm$ 0.13 & -15.81 & 1.32 x 10$^{40}$ & -0.98 \\
    124660  & 3 & 26 & 17.73 $\pm$ 0.04 & 18 & 0.73 $\pm$ 0.05 & -16.66 & 4.91 x 10$^{39}$ & -1.41 \\
    125165B & 3 & 20 & 18.26 $\pm$ 0.03 & 12 & 0.26 $\pm$ 0.03 & -16.00 & 1.57 x 10$^{39}$ & -1.91 \\
    124697  & 3 & 28 & 17.64 $\pm$ 0.02 & 16 & 0.55 $\pm$ 0.04 & -16.79 & 3.86 x 10$^{39}$ & -1.52 \\
    125180  & 3 & 20 & 17.50 $\pm$ 0.02 & 16 & 0.56 $\pm$ 0.02 & -16.77 & 4.37 x 10$^{39}$ & -1.46 \\
    125209  & 3 & 20 & 17.54 $\pm$ 0.03 & 16 & 2.20 $\pm$ 0.06 & -16.98 & 1.69 x 10$^{40}$ & -0.88 \\
    123350  & 1 & 18 & 17.71 $\pm$ 0.03 & 14 & 0.95 $\pm$ 0.05 & -12.93 & 2.03 x 10$^{38}$ & -2.79 \\
    122939  & 1 & 40 & 16.28 $\pm$ 0.03 & 22 & 1.91 $\pm$ 0.07 & -15.53 & 1.21 x 10$^{39}$ & -2.02 \\
    124040  & 3 & 18 & 18.00 $\pm$ 0.04 & 14 & 1.77 $\pm$ 0.04 & -16.06 & 1.05 x 10$^{40}$ & -1.08 \\
    123101  & 2 & 20 & 19.28 $\pm$ 0.15 & 12 & 0.01 $\pm$ 0.04 & -14.42 & … & … \\
    123213  & 4 & 24 & 17.94 $\pm$ 0.03 & 20 & 0.95 $\pm$ 0.05 & -16.86 & 9.44 x 10$^{39}$ & -1.13 \\
    123254  & 3 & 24 & 18.17 $\pm$ 0.07 & 6  & 0.12 $\pm$ 0.02 & -16.29 & 1.05 x 10$^{39}$ & -2.08 \\
    131439  & 1 & 28 & 18.08 $\pm$ 0.07 & 12 & 0.55 $\pm$ 0.05 & -10.90 & 2.57 x 10$^{37}$ & -3.69 \\
    323400  & 2 & 14 & 18.25 $\pm$ 0.04 & 8  & 0.10 $\pm$ 0.02 & -15.15 & 2.62 x 10$^{38}$ & -2.68 \\
    322496  & 1 & 28 & 17.52 $\pm$ 0.04 & 12 & 0.17 $\pm$ 0.04 & -14.03 & 8.44 x 10$^{37}$ & -3.18 \\
    323421  & 1 & 40 & 16.71 $\pm$ 0.03 & 26 & 1.18 $\pm$ 0.14 & -15.07 & 7.25 x 10$^{38}$ & -2.24 \\
    322064  & 1 & 28 & 17.46 $\pm$ 0.04 & 22 & 0.01 $\pm$ 0.01 & -14.20 & … & … \\
    322522  & 2 & 26 & 17.40 $\pm$ 0.04 & 12 & 0.09 $\pm$ 0.04 & -16.30 & 3.25 x 10$^{38}$ & -2.59 \\
    322635  & 1 & 22 & 17.52 $\pm$ 0.03 & 14 & 0.67 $\pm$ 0.04 & -13.50 & 2.04 x 10$^{38}$ & -2.79 \\
    333446  & 4 & 20 & 17.09 $\pm$ 0.02 & 12 & 1.41 $\pm$ 0.03 & -17.78 & 1.50 x 10$^{40}$ & -0.93 \\
    333330  & 4 & 30 & 17.04 $\pm$ 0.02 & 12 & 0.62 $\pm$ 0.02 & -17.73 & 5.94 x 10$^{39}$ & -1.33 \\
    334540  & 3 & 18 & 18.28 $\pm$ 0.05 & 12 & 0.24 $\pm$ 0.03 & -16.18 & 1.76 x 10$^{39}$ & -1.86 \\
    332921  & 3 & 20 & 18.81 $\pm$ 0.03 & 10 & 0.13 $\pm$ 0.02 & -15.02 & 5.15 x 10$^{38}$ & -2.39 \\
    336095  & 4 & 18 & 18.17 $\pm$ 0.02 & 18 & 2.37 $\pm$ 0.04 & -16.71 & 2.54 x 10$^{40}$ & -0.70 \\
    333337  & 1 & 24 & 17.54 $\pm$ 0.04 & 4  & 0.11 $\pm$ 0.01 & -12.85 & 1.92 x 10$^{37}$ & -3.82 \\
    336100  & 3 & 24 & 17.91 $\pm$ 0.03 & 12 & 0.02 $\pm$ 0.02 & -16.04 & … & … \\
    335430  & 2 & 18 & 18.16 $\pm$ 0.05 & 16 & 0.14 $\pm$ 0.05 & -14.63 & 2.17 x 10$^{38}$ & -2.77 \\
    334595  & 3 & 14 & 19.57 $\pm$ 0.05 & 12 & 0.08 $\pm$ 0.02 & -14.53 & 3.95 x 10$^{38}$ & -2.51 \\
    \enddata 
    \tablecomments{Measured and calculated physical parameters for all 42 observed objects. The filter number corresponds to the used narrowband filter \citep{van2016alfalfa}. The three objects with no calculated H$\alpha$ flux and SFRs were deemed non-detections in H$\alpha$.}
    \label{tab:table2}
\end{deluxetable*}

To help illustrate the nature of our sample, their derived quantities will be contrasted with observations from two comparison samples: SHIELD and the H$\alpha$ Dots survey \citep{cannon2011survey,kellar2012halpha,salzer2020halpha}. The galaxies from SHIELD were selected from primary ALFALFA observations and have \ion{H}{1} masses below 10$^{7.2}$ M$_{\odot}$.  They are, on average, more nearby and less massive than our observed galaxies \citep{mcquinn2021galaxy}. SHIELD is of particular importance to the current study due to the survey's discovery of the XMP dwarf galaxy Leoncino \citep{hirschauer2016alfalfa}. In contrast, the H$\alpha$ Dots Survey galaxies are a collection of H$\alpha$ and [O III] narrow-band detected Emission-Line Galaxies (ELGs). They were found serendipitously by searching for compact or unresolved sources of H$\alpha$ emission that were detected while primarily investigating \ion{H}{1}-detected galaxies cataloged in the ALFALFA survey \citep{van2016alfalfa}.  Only the H$\alpha$-detected H$\alpha$ Dots are included in the current comparison. The range of distances present in our current sample and the H$\alpha$ Dots are similar, although the latter catalog does reach to slightly higher redshifts.

\begin{figure}[]
    \centering
    \epsscale{1.15}
    \plotone{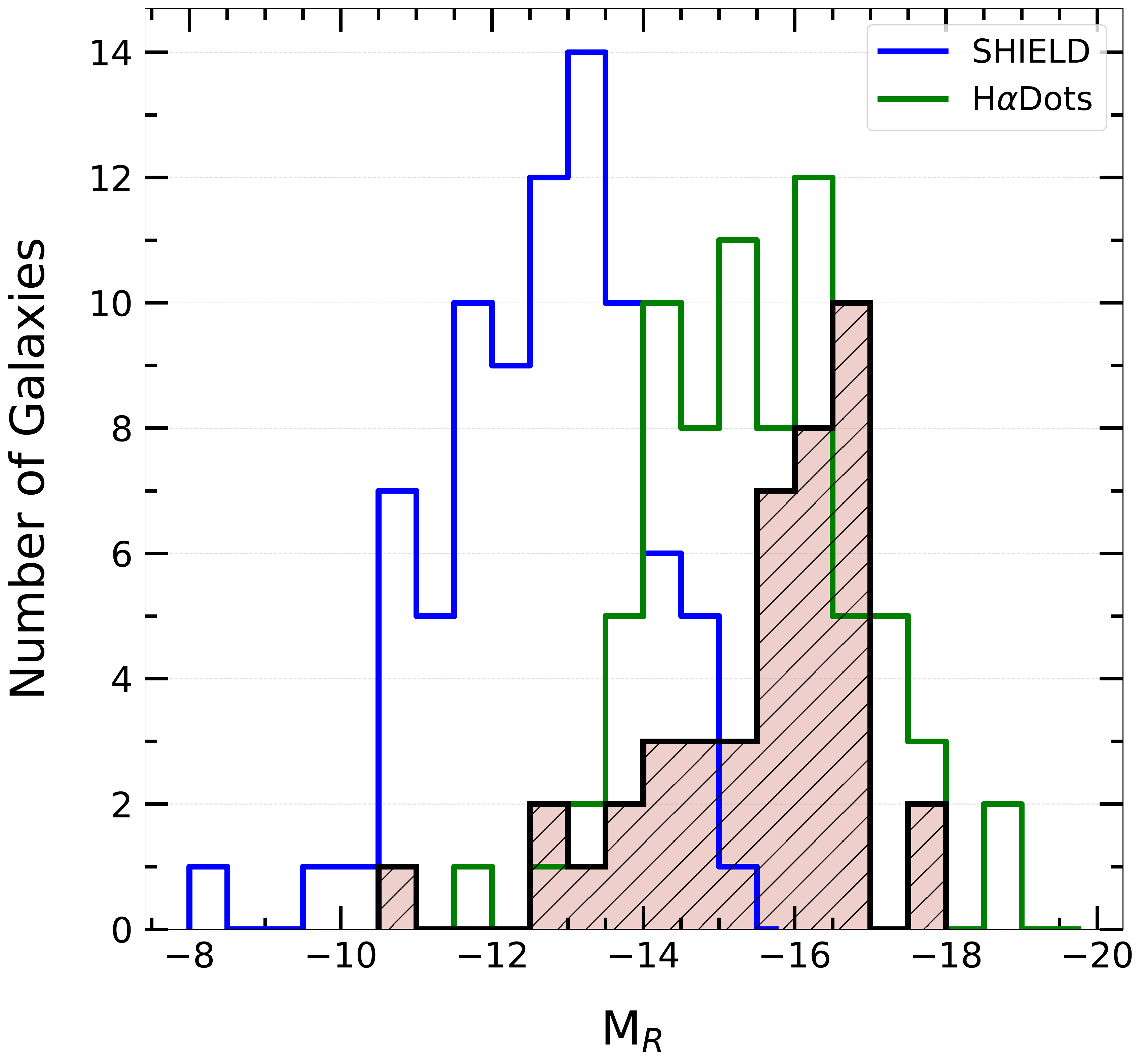}
    \caption{A histogram of the absolute magnitudes for the observed galaxies (solid) and two comparison samples: H$\alpha$ Dots (green) and SHIELD (blue). The median and average values of the M$_R$ values are $-$15.99 and $-$15.56, respectively.}
    \label{fig:MR}
\end{figure}

Figure \ref{fig:MR} shows the distribution of absolute R-band magnitudes for the observed galaxies along with the two comparison samples: SHIELD in blue and H$\alpha$ Dots in green. Our observed galaxies have absolute magnitudes that range from $-$17.78 to $-$10.90, with a median and average values of $-$15.99 and $-$15.56, respectively. For reference, the Small Magellanic Clouds (SMC) has an absolute V-band magnitude of $-$16.8 \citep{mcconnachie2012observed}, which corresponds to M$_R$ $\sim$ $-$17.3 for a characteristic dwarf galaxy color of V$-$R = 0.5.  Essentially our entire sample of ALFALFA-selected galaxies have optical luminosities less than the SMC, verifying their dwarf nature.  

The smaller and more local SHIELD galaxies are expected to have lower R-band luminosities. This expectation is borne out in Figure \ref{fig:MR}, which shows that the majority of the SHIELD galaxies have absolute magnitudes that are less luminous than the galaxies in the current sample. In contrast, the H$\alpha$ Dots, which have distances more similar to the current sample, have strong H$\alpha$ emission and R-band luminosities due to the survey's narrowband selection technique. As seen in Figure \ref{fig:MR}, the H$\alpha$ Dots have absolute R-band magnitudes similar to the current sample, with an average M$_R$ of $-$15.48, whereas the average M$_R$ for the SHIELD galaxies is $-$12.66.

\begin{figure}[]
    \centering
    \epsscale{1.15}
    \plotone{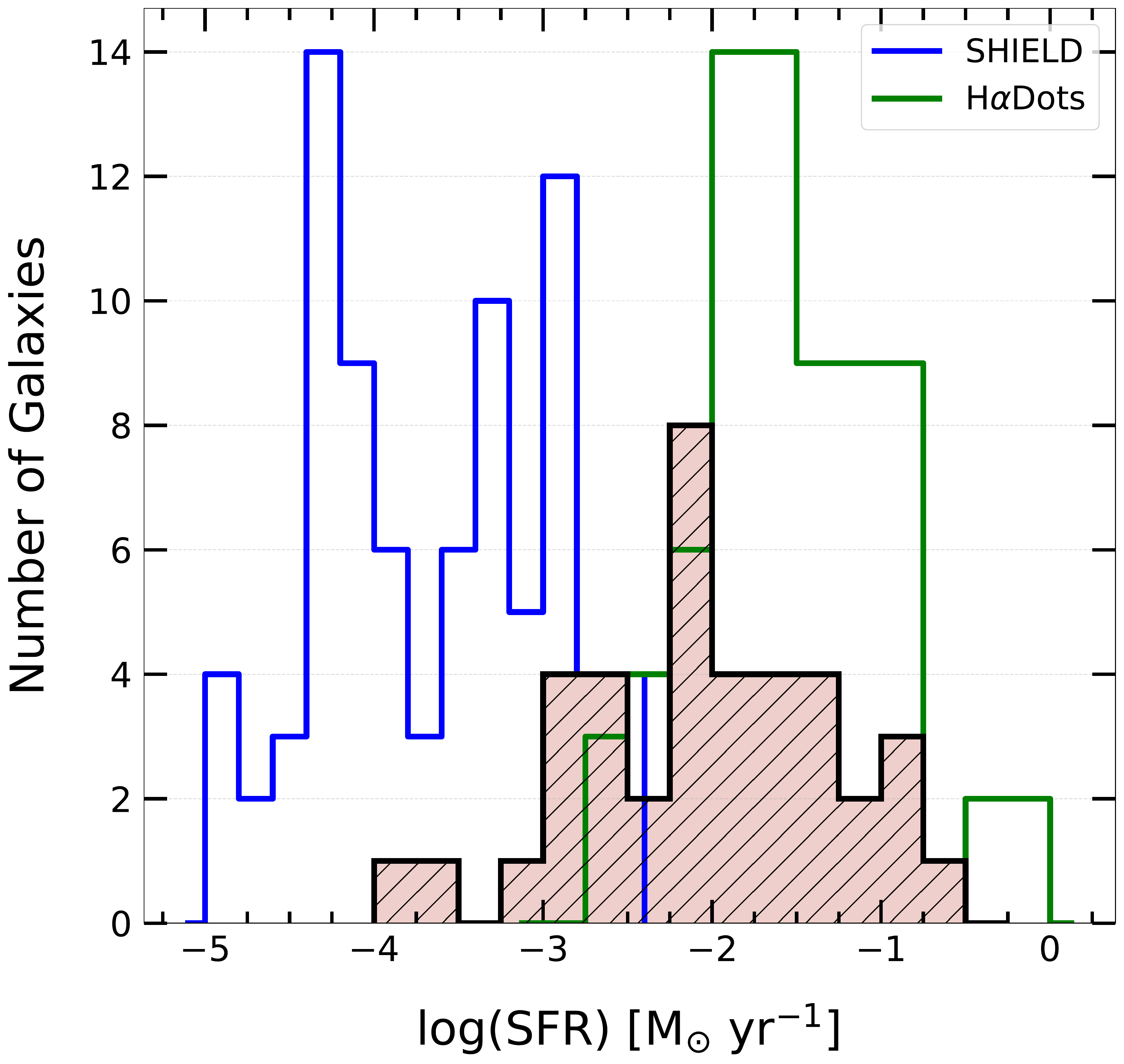}
    \caption{A histogram of the SFRs for the observed galaxies (solid) and two comparison samples: H$\alpha$ Dots (green) and SHIELD (blue). The median and average values for the log of the SFRs are $-$2.01 and $-$1.89 M$_{\odot}$ yr$^{-1}$, respectively.}
    \label{fig:SFR}
\end{figure}

Figure \ref{fig:SFR} shows the distribution of SFRs for the observed galaxies with the two comparison samples. Our observed galaxies have SFRs from 1.51 $\times$ 10$^{-4}$ to 0.20 M$_{\odot}$ yr$^{-1}$. For reference, the SFR of the Milky Way is estimated to be in the range of 0.68 to 1.45 M$_{\odot}$ yr$^{-1}$ \citep{robitaille2010present}. Therefore, the dwarf galaxy with the largest SFR, AGC 336095, has approximately one third to one fifth the SFR of the Milky Way.

Although star formation is more closely associated with cold molecular gas, neutral hydrogen is present in essentially all star-forming galaxies.  Additionally, galaxies with lower \ion{H}{1} and stellar masses will tend to have lower rates of star formation. This trend is evident with the lower \ion{H}{1} mass SHIELD galaxies, which are seen to generally have star formation rates below those in our sample. On the other hand, the H$\alpha$ Dots were selected due to their high H$\alpha$ fluxes, and thus on average have higher SFRs than the current sample. 
Overall, our current sample of ALFALFA dwarfs straddles the middle ground between the two sets of comparison galaxies, overlapping the higher end of the SFR distribution of the SHIELD sample and the low-to-mid range portion of the distribution of the H$\alpha$ Dots.

\section{Spectral Observations and Metal Abundances} \label{sec:5}

In this section we present spectroscopic observations of a subsample of our galaxies.  After describing the observations and presenting plots and measurements of the resulting spectra, we derive estimates of the metal abundances for the galaxies in the spectroscopic sample.

\subsection{Object Selection}

We constructed a sample of candidates for follow-up spectroscopy based on the outcome of our narrow-band observations.  Our sample needed to be restricted in size because we only had a limited amount of time available for the observations.  We selected the seven galaxies with the highest measured H$\alpha$ fluxes, those with F(H$\alpha$) $>$ 1.75 $\times$ 10$^{-14}$ erg s$^{-1}$ cm$^{-2}$.  While these strong-lined objects were chosen because they would yield high-quality spectra, they also tended to have intermediate-to-high luminosities, with M$_R$ values ranging from $-$14.91 to $-$16.98.  Since the goal of this project was to look for low-metallicity galaxies, we added four additional sources that had moderate H$\alpha$ fluxes (2.5 -- 9.5 $\times$ 10$^{-15}$ erg s$^{-1}$ cm$^{-2}$) but with lower R-band luminosities (M$_R$ values ranging from $-$10.90 to $-$13.60).  The choice of the latter subsample was motivated based on the well-known relationships between either  stellar mass and optical luminosity and metallicity \citep[e.g.,][]{Lequeux1979,skillman1989oxygen,berg2012,hirschauer2018metal}.  Galaxies selected for follow-up spectroscopy are listed in Table~\ref{tab:table3}.

\subsection{Observations and Emission-Line Measurements} 

Spectra were obtained for eleven galaxies using the Hobby-Eberly Telescope (HET) located at McDonald Observatory.  We employed the Low-Resolution Spectrograph 2\footnote{The Low Resolution Spectrograph 2 (LRS2) was developed and funded by the University of Texas at Austin McDonald Observatory and Department of Astronomy and by Pennsylvania State University. We thank the Leibniz-Institut für Astrophysik Potsdam (AIP) and the Institut für Astrophysik Göttingen (IAG) for their contributions to the construction of the integral field units.} (LRS2; \citealt{lrs2}), a dual-beam instrument with two separate spectral channels.  Each spectral channel is fed by independent optical fiber integral field units (IFUs).  Our observations were obtained with the Blue arm (LRS2-B), which includes the UV channel (0.496 \AA\ pixel$^{-1}$ dispersion, wavelength coverage of 3700--4700 \AA) and the orange channel (1.192 \AA\ pixel$^{-1}$ dispersion, wavelength coverage of 4600--6900 \AA).  Each galaxy was observed for 1800 s.  Only one spectrum was obtained for each source, with the exception of AGC 103425, which was observed twice.  All image processing was carried out using the Panacea pipeline software\footnote{https://github.com/grzeimann/Panacea}, with the analysis being carried out by HET staff. 

We stress that the spectra presented here are meant to be first-look measurements rather than providing a definitive spectroscopic study.  In some cases our data have yielded detection of the [\ion{O}{3}]$\lambda$4363 line, necessary for deriving the electron temperature (T$_e$) allowing for a so-called direct-method abundance.  However, in the majority of cases, our spectral data are not of sufficient quality to determine T$_e$.  In these cases we are able to estimate the metal abundance of the galaxies using one or more ``strong-line" methods.

\begin{figure}
    \centering
    \includegraphics[width=3.30in]{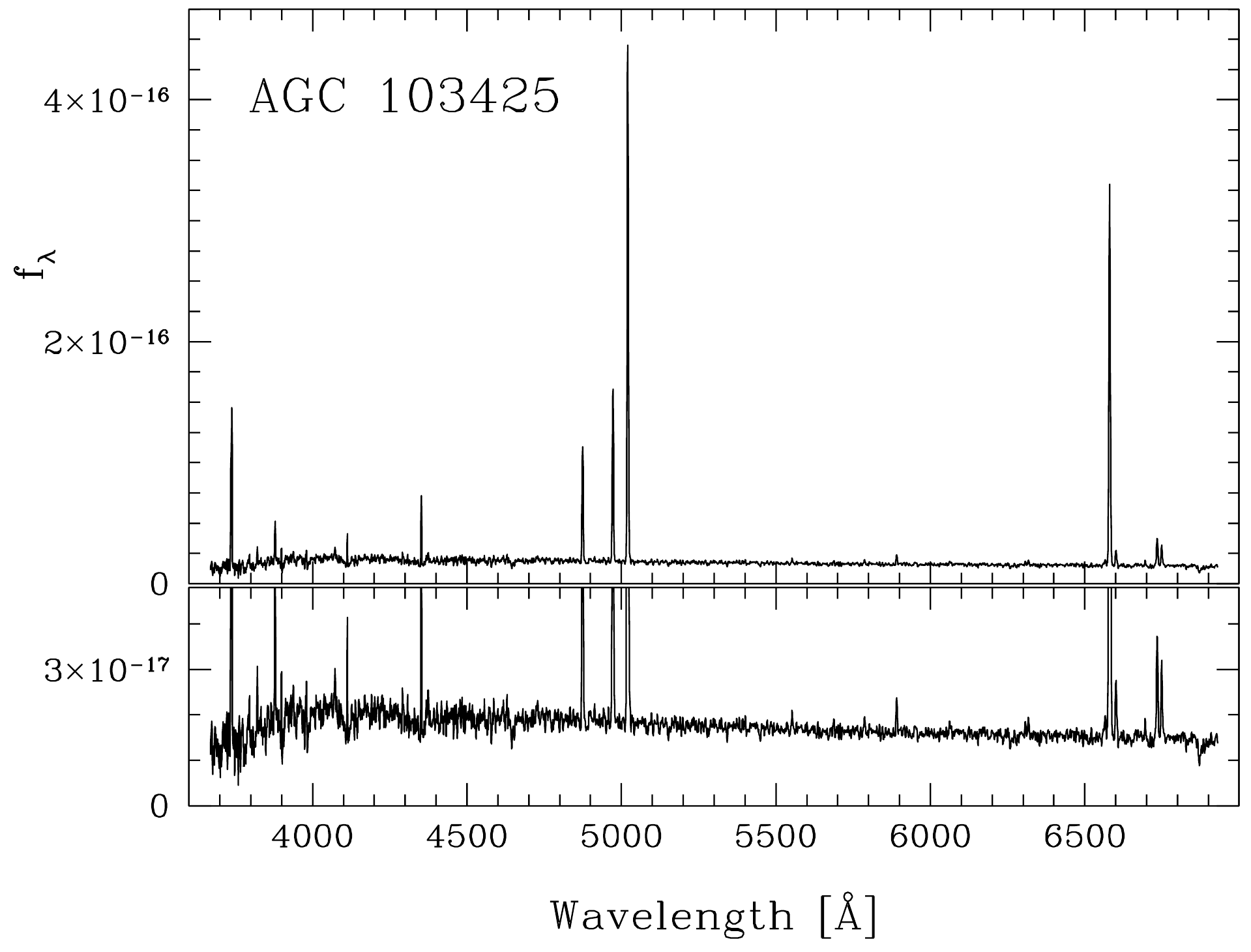}
    \includegraphics[width=3.30in]{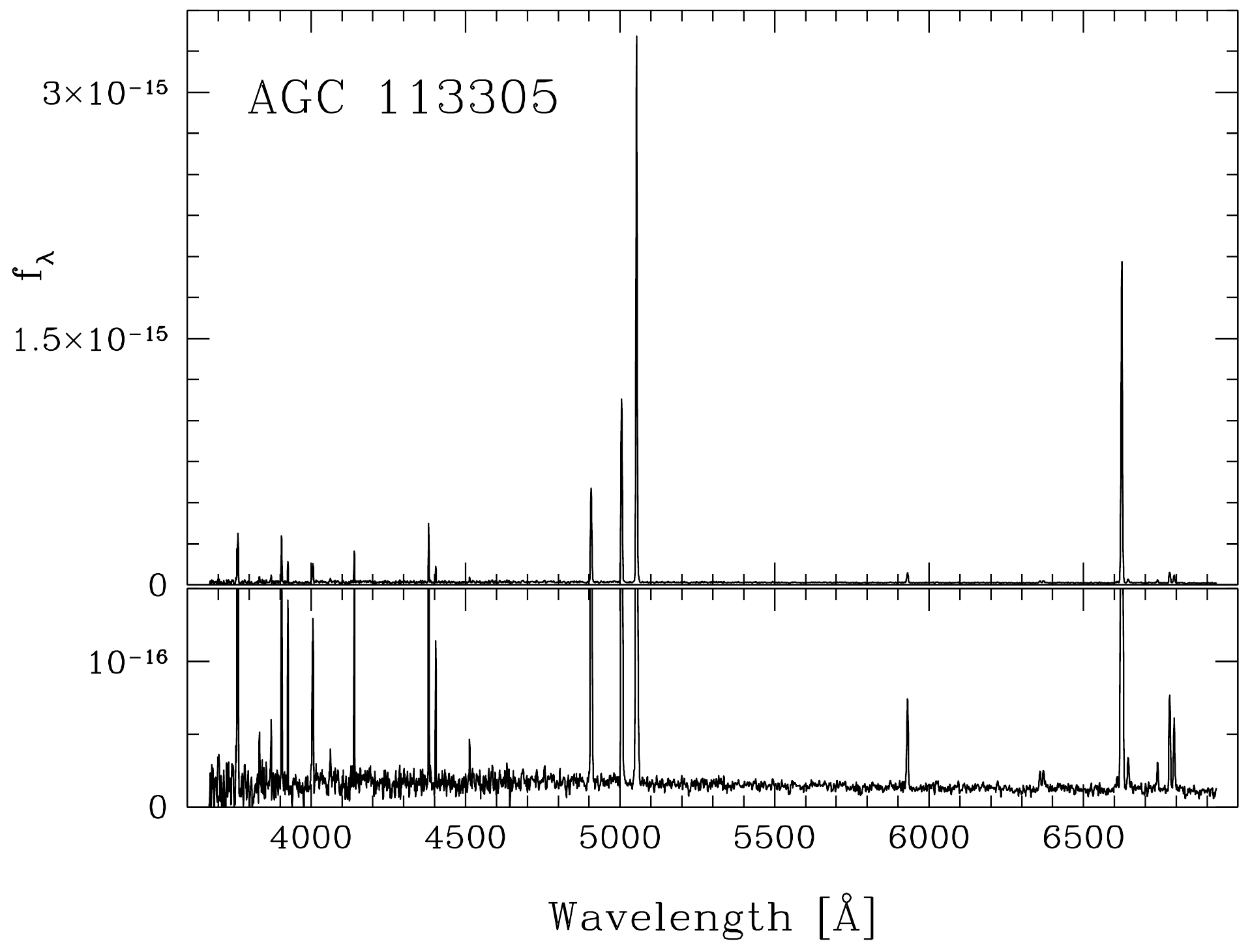}
    \includegraphics[width=3.30in]{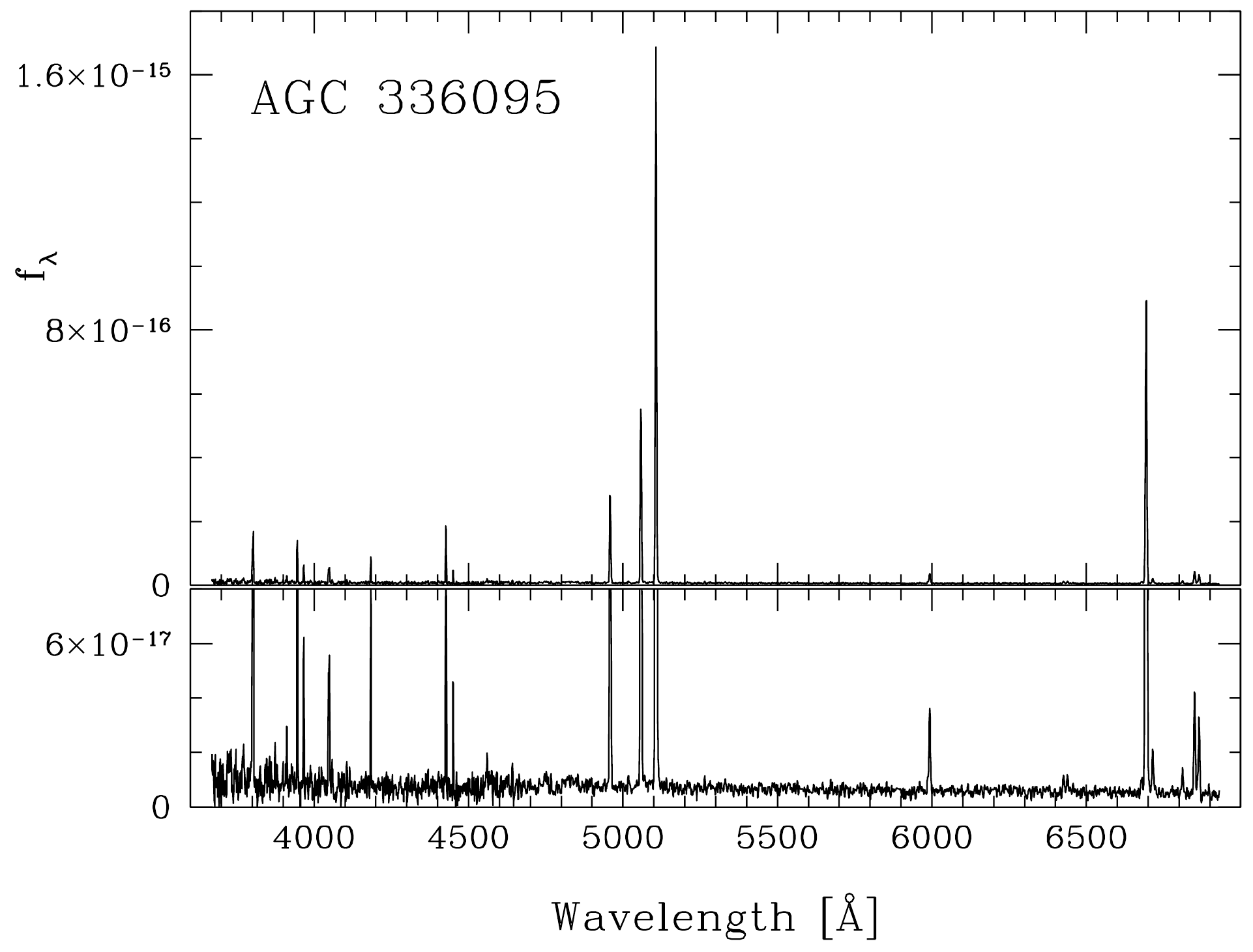}
    \caption{Spectra of three emission line regions with detected [\ion{O}{3}]$\lambda$4363 lines.  The upper section of each plot shows the full spectrum, while the lower section changes the y-axis scale to reveal the details in the weaker lines. {\it Top:} AGC 103425, one of the nearest galaxies in our sample (D = 11.7 Mpc).  
    {\it Middle:} AGC 113305 is the galaxy with the strongest H$\alpha$ flux.  {\it Bottom:} AGC 336095 is the most distant (D = 86.0 Mpc) and one of the most luminous of the galaxies in the sample (M$_R$ = $-$16.7).}
    \label{fig:te_spec}
\end{figure}

\begin{figure*}
    \centering
    \includegraphics[width=7.10in]{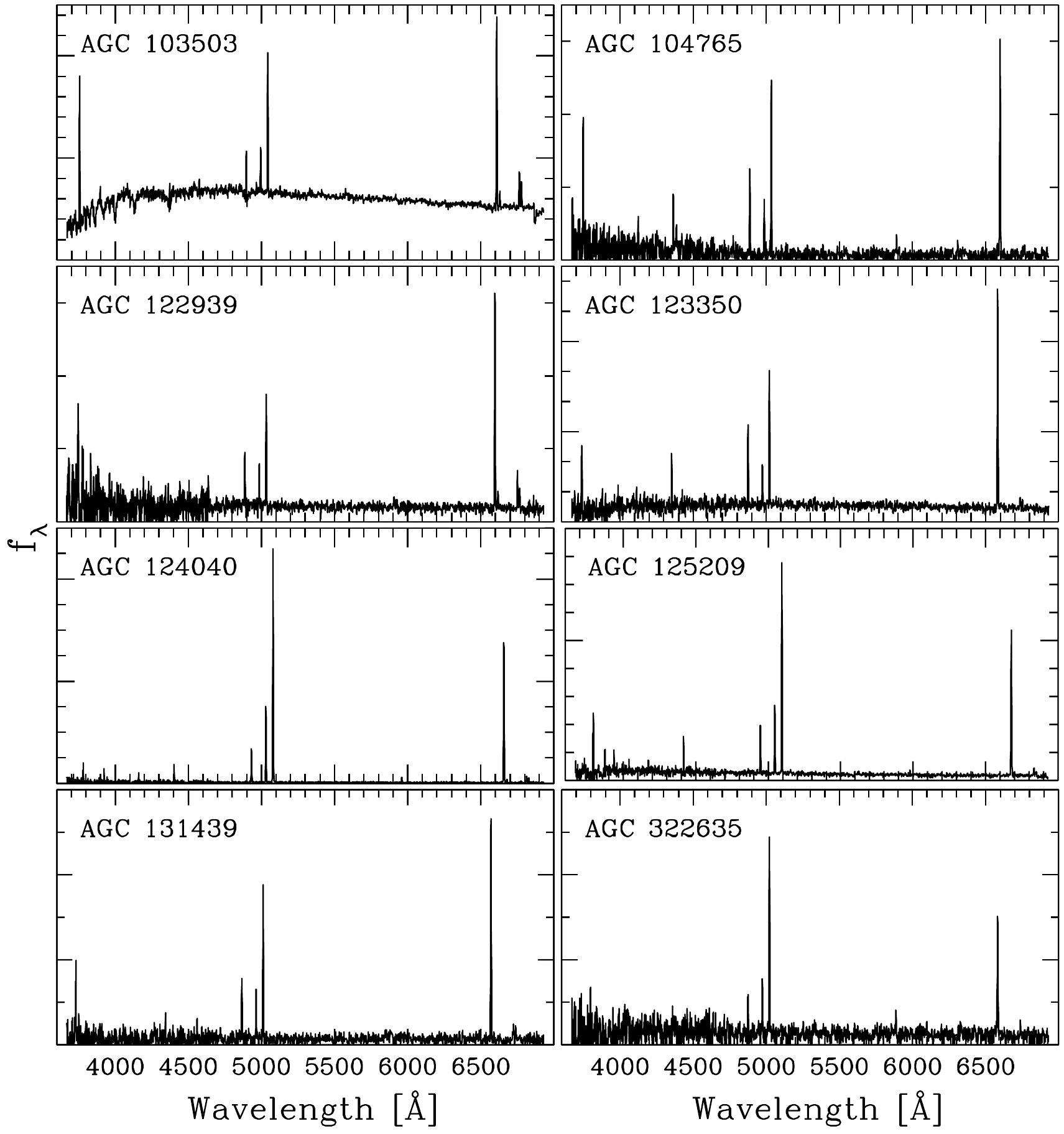}
    \caption{Spectra for the eight spectroscopically observed galaxies in our sample that did not reveal the presence of the [\ion{O}{3}]$\lambda$4363 line.  This group displays some diversity in the morphology of their spectra.  AGC 103503 shows a strong stellar continuum, while AGC 124040, AGC 125209 and AGC 322635 exhibit high-excitation spectra similar to those seen in Figure~\ref{fig:te_spec}.  AGC 104765, AGC 123350, and AGC 121439 have lower [\ion{O}{3}]/H$\beta$ ratios and weak [\ion{N}{2}] lines, characteristic of lower metallicity systems.}
    \label{fig:composite_spec}
\end{figure*}

Our spectra are presented in Figures~\ref{fig:te_spec} and~\ref{fig:composite_spec}.  The three objects shown in Figure~\ref{fig:te_spec} all exhibit a measurable [\ion{O}{3}]$\lambda$4363 line, which allows us to determine an estimate for the electron temperature T$_e$ and hence derive a direct abundance.  The full spectrum is shown in the upper portion of each spectral plot, while the lower section compresses the y-axis to allow the weaker lines to be seen in more detail.  Both AGC 113305 and AGC 336095 have very solid detections of the [\ion{O}{3}]$\lambda$4363 auroral line, while this line is only weakly detected in AGC 103425.  As seen in \S 5.3, all three yield robust T$_e$ measurements.

The spectra of eight of the galaxies did not reveal the presence of the [\ion{O}{3}] temperature sensitive auroral line, meaning that we will not be able to derive a direct abundance for them.   The spectral plots for this group are shown in Figure~\ref{fig:composite_spec}.  The eight spectra displayed reflect a diverse set of objects.  The spectrum of AGC 103503 shows a strong stellar continuum, consistent with the centralized location of the emission region.  Both this galaxy and AGC 122939 exhibit low-to-moderate [\ion{O}{3}]/H$\beta$ ratios and clear detections of the [\ion{N}{2}]$\lambda$6584 line, suggesting relatively high metal abundances.  On the other hand, three galaxies show low [\ion{O}{3}]/H$\beta$ ratios but only a very weak detection of [\ion{N}{2}]$\lambda$6584 (AGC 104765, AGC 123350, and AGC 121439), which is characteristic of galaxies with relatively low metallicities.  Finally, three galaxies have rather high-excitation spectra (large [\ion{O}{3}]/H$\beta$ ratios) and also have weak [\ion{N}{2}]$\lambda$6584 (AGC 124040, AGC 125209, and AGC 322636).  They are similar in appearance to the three galaxies plotted in Figure~\ref{fig:te_spec}, and likely have metal abundances that are intermediate between the first two groups.

We note in passing that the spectrum of AGC 104765 shows a feature at the expected location of [\ion{O}{3}]$\lambda$4363.  We did not expect to be able to detect this line in such a noisy spectrum, but measured it nonetheless.  When carried through our analysis it resulted in an estimate of the electron temperature of T$_e$ $\sim$46,000 K, which is unphysical.  We concluded that this feature is either noise, or is significantly impacted by noise, and have opted to ignore it.   Longer-exposure spectral observations of AGC 104765 might well result in a detection of [O III]$\lambda$4363 and thus a direct abundance estimate.

The reduced emission-line spectra were measured using the Python code WRALF (Wrapper for ALFA), developed by \citet{wralf}.  WRALF utilizes the ALFA code \citep{alfa}, which automatically finds emission lines in a spectrum, fits and subtracts the underlying continuum, and carries out a Gaussian fitting procress to each detected line.  The output from ALFA is a series of line positions, fluxes, equivalent widths, and line widths.  These values are used by WRALF to compute the Balmer decrement reddening coefficient and reddening-corrected emission-line ratios.  

Key emission-line ratios measured from our eleven galaxies are listed in Table~\ref{tab:table3}.  Columns 2-4 list the logarithms of the indicated emission-line flux ratios: [\ion{O}{2}]$\lambda$3727/H$\beta$, [\ion{O}{3}]$\lambda$5007/H$\beta$, and [\ion{N}{2}]$\lambda$6583/H$\alpha$.  Column 5 gives the measured value for R23, defined as R23 $\equiv$ log(([\ion{O}{3}]$\lambda\lambda$5007,4959 + [\ion{O}{2}]$\lambda$3727)/H$\beta$), while column 6 lists the value for the excitation O23 $\equiv$ log([\ion{O}{3}]$\lambda\lambda$5007,4959/[\ion{O}{2}]$\lambda$3727).

\begin{figure}
    \centering
    \includegraphics[width=3.35in]{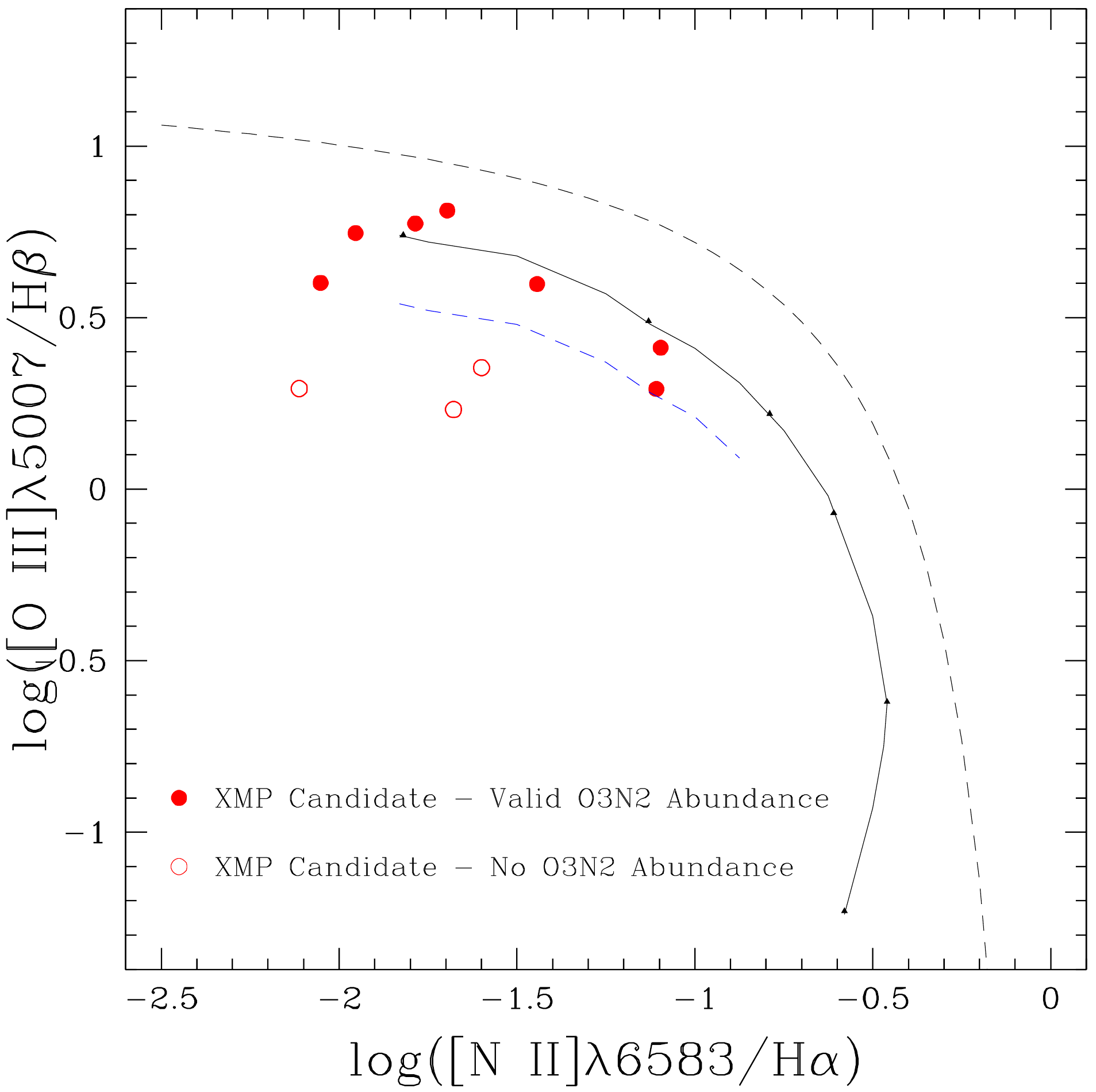}
    \caption{Emission-line ratio diagnostic diagram (a.k.a. BPT diagram) for the galaxies in our sample with spectra.  Only 10 galaxies are included because the [\ion{N}{2}] line was not detected in AGC 322635.  The dashed line is an empirical demarcation between star-forming galaxies and AGN \citep{kauffmann2003host}, while the solid black line specifies the locus of high-excitation stellar photo-ionization models from \citet{dopita1986theoretical}.   The lower blue dashed line indicates the limiting line ratios that are suitable for use in the O3N2 strong-line abundance method (see text).  All of the galaxies are located along the upper portion of the star-forming galaxy sequence, located in the regions typically occupied by intermediate- and low-metallicity systems. }
    \label{fig:BPT}
\end{figure}

We display our measured emission-line ratios in Figure~\ref{fig:BPT}, which plots the logarithm of the [\ion{N}{2}]$\lambda$6583/H$\alpha$ flux ratio {\it vs.}\ the logarithm of the [\ion{O}{3}]$\lambda$5007/H$\beta$ flux ratio (\citealt{baldwin1981classification}; hereafter BPT).  The upper dashed line in the figure is the empirical dividing line from \citet{kauffmann2003host} that separates star-forming systems from active galactic nuclei (AGN), while the solid black line denotes the locus of a series of high-excitation photo-ionization models for star-formation systems from \citet{dopita1986theoretical}.
Only ten of the eleven galaxies in our spectroscopic sample are plotted in Figure~\ref{fig:BPT} because our spectrum for AGC 322635, seen in the lower right spectrum of Figure~\ref{fig:composite_spec}, did not have a detectable [\ion{N}{2}]$\lambda$6583 line. 

The ALFALFA-selected dwarf galaxies are located in the upper-left portion of the BPT diagram.  This is the expected location for low-luminosity, low-metallicity star-forming systems.  Seven of the galaxies, denoted with filled red circles, are clustered around the high-excitation model line from \citet{dopita1986theoretical}.  This suggests that the star-formation event in these systems is relatively recent, since the higher-mass ionizing stars are still present.  The three remaining galaxies, indicated by the open circles, exhibit lower excitations indicative of older, more evolved star-forming regions.  One of these galaxies, AGC 104765, is located close to the region in the diagnostic diagram occupied by XMP galaxies \citep[e.g.,][]{izotov2012,hirschauer2016alfalfa}.  We will show in subsequent sections that the metal abundance of this galaxy is low, but not extreme.

\begin{deluxetable*}{cccccccccc}
    \tabletypesize{\small}
    \tablehead{}
    \caption{Emission Line Ratios and Abundance Measurements}
    \tablewidth{8cm}
    \tablehead{\scshape{AGC \#} & \scshape{[\ion{O}{2}]/H$\beta$} &  \scshape{[\ion{O}{3}]/H$\beta$} & \scshape{[\ion{N}{2}]/H$\alpha$} & \scshape{R23} & \scshape{O23} & \scshape{\thead{log(O/H)+12 \\ (O3N2 Method)}} & \scshape{\thead{log(O/H)+12 \\ (R23-O23 Method)}} & \scshape{\thead{log(O/H)+12 \\ (Average SLM)}} & \scshape{\thead{log(O/H)+12 \\ (T$_e$)}} \\
    (1) & (2) & (3) & (4) & (5) & (6) & (7) & (8) & (9) & (10) }
    \startdata
    103425 & \ 0.169 & 0.598 & -1.444 & 0.830 & \ 0.554 & 8.06 $\pm$ 0.15 & 7.87 $\pm$ 0.10 & 7.93 $\pm$ 0.08 & 7.93 $\pm$ 0.07 \\ 
    113305 & -0.181 & 0.746 & -1.954 & 0.908 & \ 1.052 & 7.58 $\pm$ 0.15 & 7.88 $\pm$ 0.10 & 7.78 $\pm$ 0.08 & 7.72 $\pm$ 0.03 \\
    336095 & -0.134 & 0.774 & -1.786 & 0.937 & \ 1.033 & 7.69 $\pm$ 0.15 & 7.95 $\pm$ 0.10 & 7.87 $\pm$ 0.08 & 7.86 $\pm$ 0.07 \\
    103503 & \ 0.591 & 0.412 & -1.096 & 0.866 & -0.054 & 8.37 $\pm$ 0.15 & 8.52 $\pm$ 0.10 &  8.47 $\pm$ 0.08 & ... \\ 
    104765 & \ 0.112 & 0.293 & -2.112 & 0.593 & \ 0.306 & ... & 7.60 $\pm$ 0.10 & 7.60 $\pm$ 0.10 & ... \\
    122939 & \ 0.343 & 0.292 & -1.109 & 0.682 & \ 0.074 & 8.41 $\pm$ 0.15 & 8.78 $\pm$ 0.10 & 8.68 $\pm$ 0.08 & ... \\
    123350 &  -0.032 & 0.232 & -1.678 & 0.506 & \ 0.389 & ... & 7.46 $\pm$ 0.10 & 7.46 $\pm$ 0.10 & ... \\
    124040 & -0.112 & 0.812 & -1.696 & 0.974 & \ 1.049 & 7.73 $\pm$ 0.15 & 8.03 $\pm$ 0.10 & 7.94 $\pm$ 0.08 & ... \\ 
    125209 & \ 0.123 & 0.601 & -2.052 & 0.823 & \ 0.603 & 7.62 $\pm$ 0.15 & 7.85 $\pm$ 0.10 & 7.78 $\pm$ 0.08 & ... \\ 
    131439 & \ 0.165 & 0.354 & -1.600 & 0.650 & \ 0.314 & ... & 7.67 $\pm$ 0.10 & 7.67 $\pm$ 0.10 & ... \\
    322635 & ... & 0.647 & ... & ... & ... & ... & ... & ... & ... \\
    \enddata 
    \tablecomments{Emission line ratios and abundance measurements for the eleven galaxies with spectra. All line ratios in columns 2-6 are logarithmic values. [\ion{N}{2}]/H$\alpha$ \& [\ion{O}{3}]/H$\beta$ are emission line ratios used to create the standard emission line diagnostic diagram in Figure \ref{fig:BPT}. R23 \& O23 are defined in the text and used to create Figure \ref{fig:Mgh}. Strong-line method (SLM) abundances derived using the O3N2 and R23-O23 methods are listed in columns 7 and 8, respectively.  Average SLM abundances are given in column 9, while the direct abundance measurements for the three galaxies with the [\ion{O}{3}]$\lambda$4363 temperature-sensitive line are given in column 10.}
    \label{tab:table3}
\end{deluxetable*}

\subsection{Abundance Measurements}

With the exception of AGC 322635, the spectra of the galaxies presented in the previous section are suitable for deriving metallicity estimates.  Our procedures for measuring their abundances follow standard practices.  First we describe the so-called strong-line method process, after which we present the T$_e$, or direct method, procedure for the three galaxies with detected [\ion{O}{3}]$\lambda$4363.

\subsubsection{Strong-line Methods}

There are numerous examples of calibrations between oxygen metal abundance and commonly observed strong emission-line ratios.  The current analysis utilizes two of these strong-line methods.

\begin{figure}
    \centering
    \includegraphics[width=3.35in]{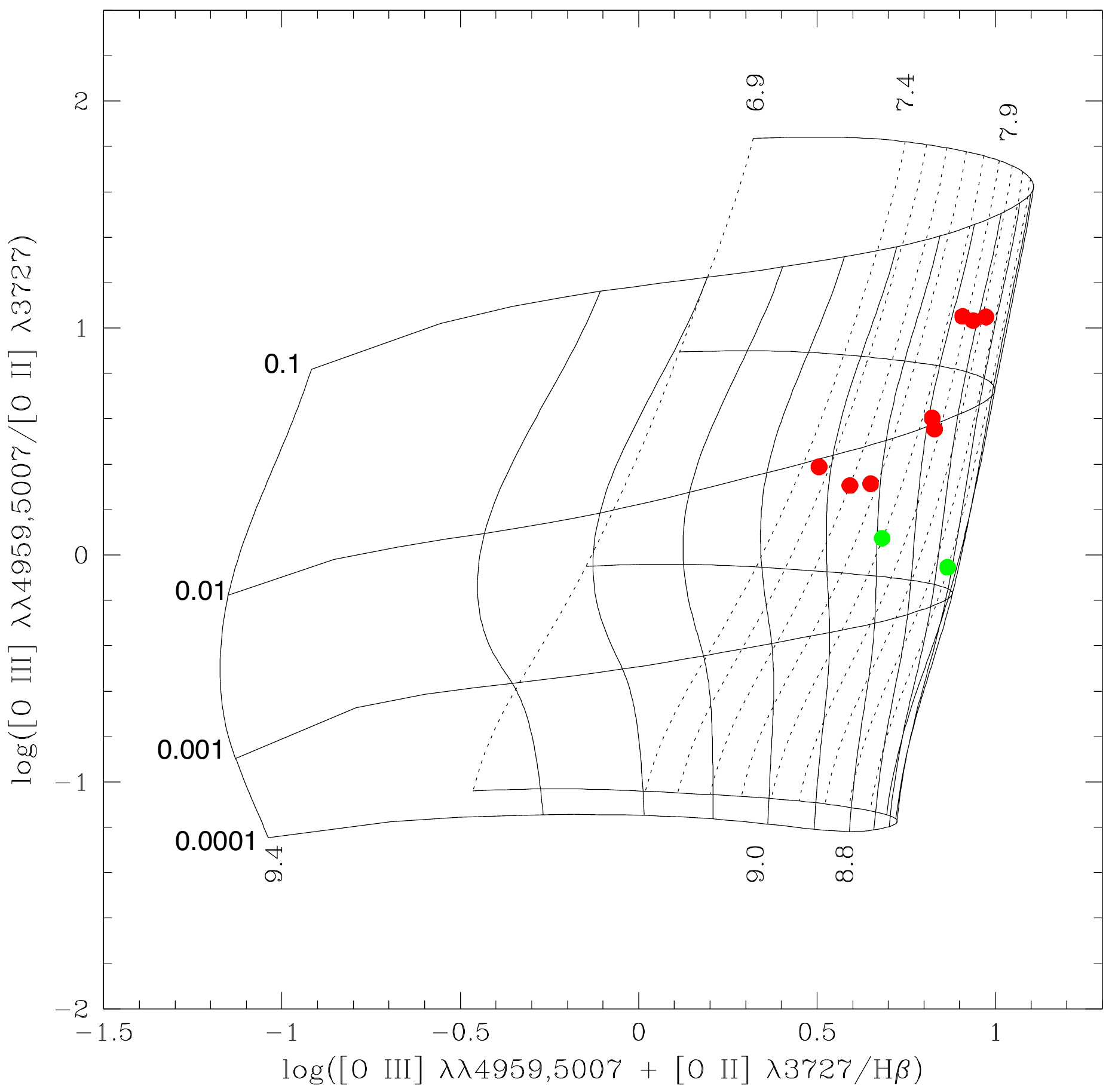}
    \caption{R23-O23 diagram for our sample of ALFALFA-selected dwarf galaxies.  The model grid is from \citet[][60 M$_\odot$ upper mass limit]{mcgaugh1991h}.  The numerical labels on the left side of the grid indicate the ionization potentials used in the models, while the labels along the top and bottom curves indicate the metal abundance (log(O/H)+12).  Green circles represent our galaxies that are located on the front side of the grid (higher abundances) while the red circles are located on the lower metallicity far side of the grid. }
    \label{fig:Mgh}
\end{figure}

\paragraph{O3N2 Method} The method was first introduced by \citet{alloin1979}, and utilizes the O3N2 emission-line diagnostic that is created by combining the two line ratios plotted in the BPT diagram (O3N2 $\equiv$ log(([\ion{O}{3}]$\lambda$5007/H$\beta$)/([\ion{N}{2}]$\lambda$6583/H$\alpha$))).  We adopt the O3N2 calibration presented in \citet{hirschauer2018metal} to generate estimates of the O/H abundance for our galaxies.  The resulting metallicities are listed in column 7 of Table~\ref{tab:table3}.  The RMS scatter in the O3N2-metallicity calibration in \citet{hirschauer2018metal} is 0.11 dex in O/H; we adopt a more conservative estimate of 0.15 dex for the uncertainties in the abundances derived using this method.

We note that the \citet{hirschauer2018metal} O3N2 calibration does not yield accurate results when applied to galaxies whose emission-line ratios lie outside of the region occupied by the calibrating galaxies.  The blue dashed line in Figure~\ref{fig:BPT} represents the lower limit for objects in the BPT plot to have their abundances reliably derived with this method.  The three ALFALFA dwarfs plotted as open circles in Figure~\ref{fig:BPT} are located below this boundary and hence do not have O3N2 abundances listed in Table~\ref{tab:table3}.

\paragraph{R23-O23 Method} This method is a variant on the R23 method first proposed by \citet{pagel1979}, where the quantity R23 is defined above in the previous subsection.  The R23 method has been utilized in numerous studies \citep[e.g.,][]{edmunds1984,mccall1985,skillman1989oxygen,zaritsky1994,pilyugin2001,pettini2004,pilyugin2005,hirschauer2015}, but suffers from the fact that the correlation between R23 and O/H is double valued. \citet{mcgaugh1991h} extended the R23 method by introducing the additional quantity O23 (sometimes referred to as the excitation).  He utilized CLOUDY \citep{cloudy,cloudy17} to generate models of stellar photo-ionized nebulae and produced grids in R23-O23 space that allow for the estimation of the metallicity of objects based on these two strong-line ratios.  

We plot O23 {\it vs.} R23 for our ALFALFA-selected dwarf galaxies in Figure~\ref{fig:Mgh}, using the values tabulated in Table~\ref{tab:table3}.  We over-plot a model grid from \citet{mcgaugh1991h} (upper mass limit of 60 M$_\odot$).  Using the combination of the measured line ratios and the model grid, we derive metallicity estimates for the ten galaxies in our sample with the necessary lines.  In order to resolve the ambiguity of which portion of the model grid each galaxy is located, we utilize the additional line ratio [\ion{O}{2}]/[\ion{N}{2}], which has been shown to be effective at distinguishing which branch of the R23 relation a galaxy is on \citep{vanzee1998}.  In the case of the current sample, there is very little confusion on this issue.  Two galaxies in our sample appear to be located in the higher metallicity branch of the relation (i.e., the ``front side" of the grid shown in Figure~\ref{fig:Mgh}); these objects are plotted as green circles.  The remaining objects are all located on the lower metallicity branch (i.e., the ``back side" of the model grid).  These latter objects are plotted as red circles.  The abundances derived using the R23-O23 method are listed in column 8 of Table~\ref{tab:table3}. We adopt uncertainties of 0.10 dex as suggested by  \citet{mcgaugh1991h} for abundances derived using this method.

\paragraph{Final Strong-Line Abundances} Since our metallicity estimates using the two strong-line methods described above are largely independent of each other, we adopt the strategy of averaging the results and using the mean value as our final estimate of the strong-line abundance.  These (weighted) average values are listed in column 9 of Table~\ref{tab:table3}.  For the three galaxies without O3N2 method abundances we simply list the R23-O23 method values in this column.  We note that for the three galaxies with direct abundance measurements (next section), the average strong-line abundances agree very well with the derived direct-method metallicities

\subsubsection{Direct Method}

For the three sources illustrated in Figure~\ref{fig:te_spec}, we were able to detect and measure the temperature sensitive [\ion{O}{3}] auroral line located at 4363 \AA.  This in turn allowed us to derive an estimate of the electron temperature T$_e$ for each source.  Having access to a reliable measurement for T$_e$ means that the oxygen abundance can be computed directly \citep[e.g.,][]{AGN2}.  We utilized the Emission Line Spectrum Analyzer (ELSA) code \citep{ELSA} to carry out the abundance analysis.   

Our methodology follows procedures described in \citet{hirschauer2022}.  In brief, we utilize the measured emission-line fluxes to infer the electron temperatures and densities of the sources.   These are used by ELSA to compute the emissivities for each collisionally-excited metal-atom line, from which the ionic abundances are derived.  Following standard practice, we define the relative oxygen abundance O/H as being equal to the sum of the singly and doubly ionized oxygen abundances: (O$^+$ + O$^{++}$)/H$^+$.  Our final direct-method abundances for these three ALFALFA dwarfs are listed in column 10 of Table~\ref{tab:table3}.  The uncertainties listed are derived by utilizing the formal uncertainties in the electron temperatures and computing the ionic abundances at T$_e$ $\pm$ $\sigma_{T_e}$.

As seen in Table~\ref{tab:table3}, our direct-method abundances are in good agreement with the average strong-line abundances for the three dwarf galaxies with both.  AGC 113305, the ALFALFA dwarf with the strongest emission-line flux from our narrow-band imaging data, yields the most accurate direct abundance (uncertainty of 0.03 dex) and is seen to have an O/H abundance ratio of $\sim$1/10th of the solar value, adopting (log(O/H)+12)$_\odot$ = 8.69 \citep{asplund2009}.  For AGC 103425 and AGC 336095, the direct abundances have somewhat higher uncertainties (0.07 dex), which is a reflection of the somewhat higher uncertainties in the derived temperatures (12830 $\pm$ 731 K and 14500 $\pm$ 760 K, respectively).


\section{Discussion} \label{sec:6}

As stated in the introduction of this paper, a key goal of this study is to test the efficacy of using \ion{H}{1}-selected galaxy samples to detect XMP galaxies.  The current approach tests a specific sample-selection method described in \S 2.  As documented in Table~\ref{tab:table3}, we have not succeeded in discovering any new XMP galaxies.  The median O/H value of our sample is 7.82 (13\% solar).  However, there are three galaxies with abundances below 10\% solar.  The lowest metallicity measured for this sample is 7.46 (6\% solar), for AGC 123350.   Hence, while we have failed to discover any new XMP galaxies, we did find a number of systems with relatively low abundances.

The lack of success in finding new XMP galaxies is perhaps not a major surprise.  The original sample we selected was based purely on the \ion{H}{1} characteristics of the galaxies.  In particular, we utilized the lower S/N ALFALFA galaxies with the idea that, at any given distance, they would have lower \ion{H}{1} masses than their main ALFALFA catalog counterparts.  In addition, they were selected to be located in the low-density region in front of the Pisces-Perseus supercluster \citep[e.g.,][]{hg1986}, because some studies have suggested that void galaxies tend to preferentially have lower metal abundances \citep[e.g.,][but also see \citealt{wegner2019}]{pustilnik2016,kniasev2018}.

Despite using these selection criteria that are geared toward finding low-metallicity sources, the final sample can be seen to include a broad range of galaxy characteristics.  Figure~\ref{fig:MHI} reveals that while our sample of ALFALFA dwarfs includes several low-mass galaxies, the bulk of the sample has \ion{H}{1} masses between 10$^8$ and 10$^9$ M$_\odot$.  Since dwarf galaxies exhibit a broad-range of \ion{H}{1}-to-stellar masses, we could not immediately dismiss these higher \ion{H}{1} mass objects as being irrelevant.  Nonetheless, Figure~\ref{fig:MHI} hints at the fact that much of the sample will likely not be good XMP candidates.

The optical characteristics of the ALFALFA dwarfs, as illustrated in Figures~\ref{fig:MR} and \ref{fig:SFR}, further re-enforce the broad range of galaxian properties present in the sample.  While the R-band absolute magnitudes do indicate that these are indeed dwarf galaxies (M$_R$ $>$ $-$17), the majority are more luminous than M$_R$ = $-$14.  The ALFALFA sample overlaps the luminosity distribution of the low \ion{H}{1} mass SHIELD sample, but has a mean M$_R$ value that is near the middle of the higher-mass, higher-luminosity H$\alpha$ Dots.  The latter sample has recently been shown to have a metallicity distribution quite similar to that of the current sample \citep{hirschauer2022}.   Similarly, the distribution of SFRs shown in Figure~\ref{fig:SFR} shows that the ALFALFA dwarfs in the current sample do have a modest overlap with the SHIELD galaxies, indicative of low-mass systems.  However, the majority of our sample have higher SFRs and are more similar to the H$\alpha$ Dots sample.

\begin{figure}
    \centering
    \includegraphics[width=3.35in]{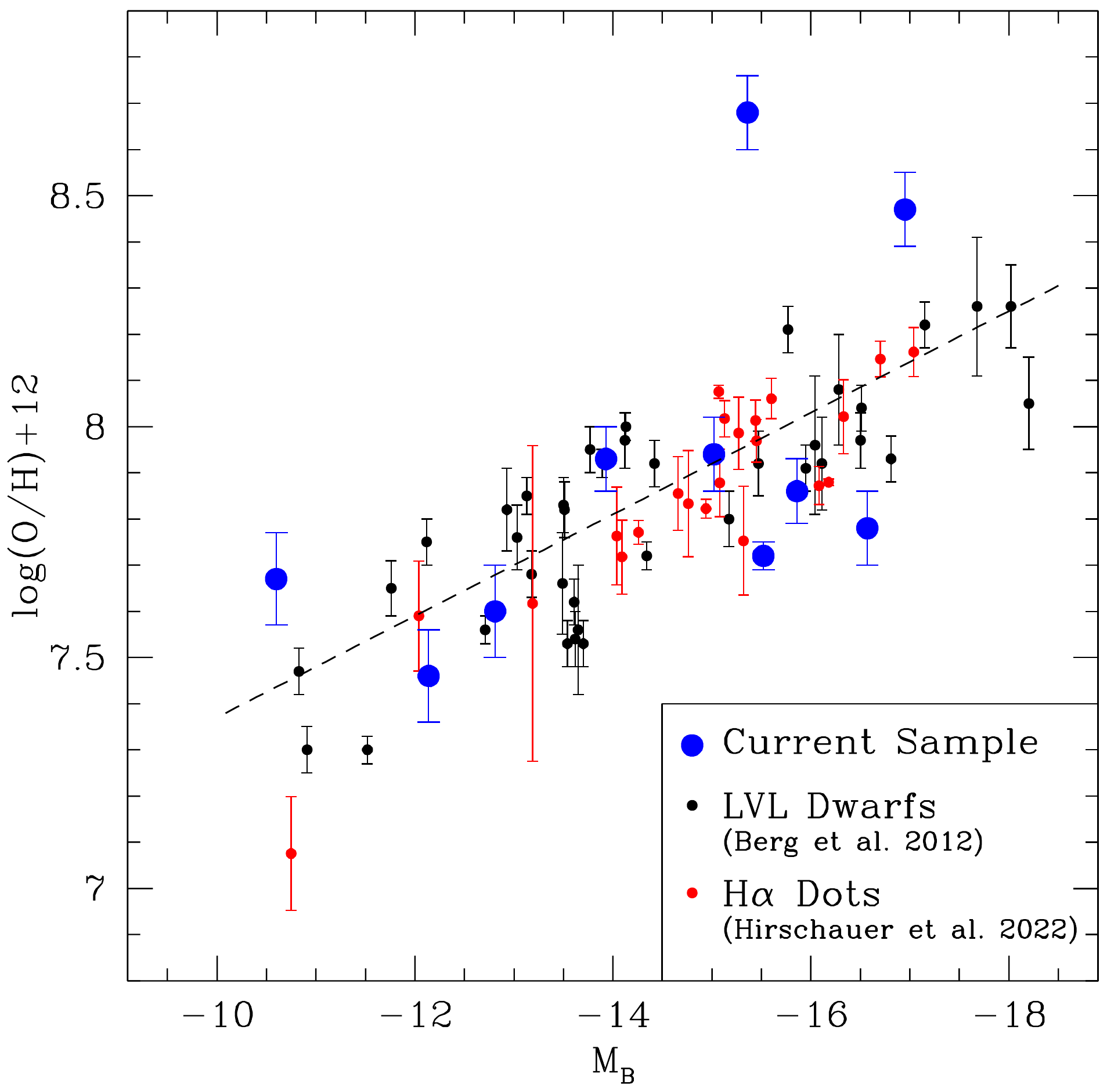}
    \caption{Luminosity-metallicity relation (LZR) plot for the current sample.  The ALFALFA dwarfs are plotted as blue circles.   Also shown, for comparison, are the \citet{berg2012} LVL sample (black dots) and the \citet{hirschauer2022} H$\alpha$ Dots sample (red dots). The dashed line is the formal linear fit to the LVL sample, also from \citet{berg2012}.  The ALFALFA dwarfs are seen to follow the same LZR defined by the other dwarf galaxy samples.  The most extreme outlier from the current sample, AGC 122939, is discussed in the text.}
    \label{fig:LZR}
\end{figure}

To further illustrate these points, we plot the ALFALFA dwarfs in a luminosity-metallicity relation (LZR) plot in Figure~\ref{fig:LZR}.  For comparion, we also plot the nearby dwarf irregular galaxy sample from the Local Volume Legacy (LVL) survey (\citealt{berg2012}; black dots) and H$\alpha$ Dots with direct abundances from \citeauthor{hirschauer2022} (\citeyear{hirschauer2022}; red dots).  Galaxies in the current sample are plotted as blue circles.  To facilitate this comparison, we converted our R-band absolute magnitudes to B-band using a procedure similar to the one described in \citet{hirschauer2022}.  We note in passing that the H$\alpha$ Dot galaxy with the lowest metal abundance in Figure~\ref{fig:LZR} is in fact AGC 198691, a.k.a. the Leoncino dwarf.  This XMP dwarf was independently detected in the H$\alpha$ Dots survey after having been included in the \ion{H}{1}-selected SHIELD survey \citep{cannon2011survey,mcquinn2020leoncino}.  

As is evident in Figure~\ref{fig:LZR}, the ALFALFA dwarfs tend to follow the same LZR that is defined by the LVL and H$\alpha$ Dot dwarfs.  With a few exceptions, the ALFALFA dwarfs are located within the envelope of points defined by the galaxies from these other samples in the luminosity-metallicity plane.  This result supports the picture that emerges from the discussion above that the current sample of \ion{H}{1}-selected dwarfs are not particularly exceptional with regard to their optical characteristics or metallicities.  Rather, they have properties that place them squarely in the domain filled by optically-selected dwarf galaxies.

\subsection{Outliers in the LZR: A possible tidal dwarf galaxy?}

Two of the ALFALFA dwarfs are located above the main locus of points in the LZR diagram.  AGC 103503 (M$_B$ = $-$16.95, log(O/H)+12 = 8.47) is only slightly elevated in abundance relative to the comparison samples.  If one adopts the O3N2 abundance (log(O/H)+12 = 8.37 $\pm$ 0.15) rather than the weighted mean of the two strong-line abundances, it would be located within the envelope of values from the comparison data set.  The more extreme outlier in Figure~\ref{fig:LZR} is AGC 122939 (M$_B$ = $-$15.36, log(O/H)+12 = 8.68), which is located more than 0.7 dex above the trend line from \citet{berg2012}.  As with AGC 103503, its weighted mean strong-line abundance is perhaps inflated by a higher R23-O23 value (8.78 $\pm$ 0.10).  However, its O3N2 abundance of 8.41 $\pm$ 0.15 is still more than 0.4 dex above the trend line.

Examination of the spectrum of AGC 122939 (Figure~\ref{fig:composite_spec}) reveals an extremely noisy blue end.  This may motivate questions regarding the integrity of our [\ion{O}{2}]$\lambda$3727 measurement, which might cause our R23-O23 abundance estimate to be off.  However, the nebular lines in the red portion of the spectrum ($\lambda$ $>$ 4700 \AA) are well defined and clearly indicate an abundance in the range specified by our strong-line method analysis.   Hence, this galaxy does appear to be genuinely metal rich for its luminosity.  This raises the possibility that this object was created as a tidal dwarf \citep[e.g.,][]{duc2012,gray2022}.  

Tidal dwarfs are created during mergers or close encounters between two larger galaxies, when gas and some existing stars are pulled from one of the systems and end up coalescing into a new, self-gravitating dwarf galaxy.  A tell-tale sign that a dwarf galaxy was created via this mechanism is that they tend to exhibit higher-than expected abundances for their mass and luminosity.  This is because the parent galaxy is typically much more massive than the tidal dwarf, and will have substantially higher abundances.  The tidal dwarf forms from the enriched gas removed from disk of the parent galaxy.  AGC 122939 exhibits this type of enrichment.  

We explored the possibility that AGC 122939 could be of a tidal origin by considering its local environment.  Utilizing the deep redshift catalog used to create Figure~\ref{fig:cone}, we identified all galaxies with projected (minimum) separations of less than 1.2 Mpc and velocity offsets of less than 600 km s$^{-1}$.   For reference, AGC 122939 is the green circle in Figure~\ref{fig:cone} located just below the top boundary of the cone with a velocity of 1523 km s$^{-1}$.  Four galaxies were found that satisfied the proximity criteria above.  Three are dwarf galaxies with \ion{H}{1} masses and optical luminosities comparable to AGC 122939.  This makes them unlikely to be the parent galaxy.  The fourth galaxy is UGC 1970, an edge-on spiral with an \ion{H}{1} mass 15-20 times larger than AGC 122939.  UGC 1970 has the smallest projected (minimum) separation of any galaxy from AGC 122939 of 550 kpc.   The velocity difference between the two galaxies is 390 km s$^{-1}$.

The current separation between the two galaxies is fairly large, and neither galaxy exhibits any indication of tidal distortions in their archival optical images.  We can derive an order of magnitude estimate for the amount of time that has elapsed since the tidal encounter that created AGC 122939, assuming UGC 1970 is in fact the parent galaxy.  First we assume that the two galaxies are at approximately the same distance, so that the projected separation is similar to their true separation.   Then we adopt the approximation that the observed velocity difference between the two galaxies is comparable to how fast the two galaxies are receding from each other in the plane of the sky.  This results in a minimum time since the tidal encounter of $\sim$1.4 Gyr.   Such a long time since the creation of the tidal dwarf is consistent with the lack of any remaining visible indication of tidal debris.

We conclude, based on the available evidence, that we cannot rule out the possibility that AGC 122939 was formed from a tidal encounter with UGC 1970 between 1 and 2 Gyr in the past.  However, we do not consider this scenario to be very likely.  For one thing, we do not find another massive galaxy within the vicinity of the AGC 122939 - UGC 1970 pair that could have triggered the creation of the tidal dwarf.  For the time being, we are unable to propose a robust explanation for the elevated metallicity observed in AGC 122939.  While the tidal dwarf scenario remains viable, we do not feel that there is sufficient evidence to adopt it as the preferred explanation.

An alternate explanation is that AGC 122939 and, to a lesser extent AGC 103503, have inflated N/O ratios which lead to an over-estimate of their strong-line abundances, similar to the galaxies studied by \citet{berg2011}.  However, in the \citet{berg2011} study the high N/O systems revealed more normal abundances when the strong-lined method used did not include use of the [\ion{N}{2}] line (e.g., R23-O23).   In the case of both AGC 122939 and AGC 103503, their R23-O23 abundances are {\it higher} than their O3N2 abundances.   Furthermore, neither galaxy has a high [\ion{N}{2}]/[\ion{O}{2}] ratio.  Hence, we do not believe that these galaxies are exhibiting higher O/H values because of high N/O ratios.

\subsection{An Alternative Approach to using \ion{H}{1}-selected samples to search for XMP galaxies}

The fact that our current study has failed to detect any new XMP galaxies does not necessarily imply that using \ion{H}{1}-selected galaxy samples to search for low-metallicity objects is a bad idea.  Rather, it implies that the specific approach to constructing our sample was flawed.  While the approach had merit, in the end it did not yield the desired results. But are there alternative selection schemes that might be employed that could improve the chances of finding new XMP galaxies?

One fairly obvious extension of the current \ion{H}{1} selection procedure would be to combine both optical and 21-cm data to better focus in on the lowest stellar {\it and} \ion{H}{1} mass objects.  This is effectively what is done with the SHIELD sample \citep{cannon2011survey}.  The 82 galaxies that make up the full SHIELD sample were selected to have some of the lowest \ion{H}{1} masses detected by ALFALFA (M$_{HI}$ $\leq$ 10$^{7.2}$ M$_\odot$).  In addition, SHIELD galaxies must include a detectable optical counterpart.  Many of these are galaxies with very low optical luminosities and stellar masses.    As seen in Figure~\ref{fig:MR}, there are 25 SHIELD galaxies with M$_R$ less luminous than $-$12, compared to only one galaxy each from the current sample and the H$\alpha$ Dots.  Obtaining metallicity estimates for the lower luminosity SHIELD galaxies would appear to be fertile ground in the search for new XMP galaxies.  Members of the SHIELD team are currently actively engaged in acquiring spectra for the SHIELD galaxies, with the hope of uncovering additional XMP systems similar to the Leoncino dwarf (J. Salzer, Private communication).

\section{Summary and Conclusion} \label{sec:7}

After the discovery of two of the most metal-poor objects known in the ALFALFA-based galaxy samples of the SHIELD and UCHVC projects, additional methods for selecting potential XMP candidates from the ALFALFA catalogs are being explored. The current study presents a metallicity investigation based on a sample of 78 ALFALFA-selected dwarf galaxies chosen for their low \ion{H}{1} signal-to-noise ratios (4.0 - 6.5), low redshifts, and, in some cases, their low-density environments. Of the 78 selected galaxies, 43 galaxies were observed with narrowband observations to measure their current star-formation rates. Spectra were obtained for the seven galaxies with the highest measured H$\alpha$ fluxes, and for four additional galaxies with moderate H$\alpha$ fluxes but lower \ion{H}{1} masses and optical luminosities.  Metallicity measurements were derived for ten of the eleven galaxies with spectra.

These ten galaxies span $\sim$1.2 dex in metallicity from log(O/H)+12 = 7.46 to 8.68, with a median value of 7.82.  None of the galaxies in our current sample have abundances low enough to be classified as XMP galaxies (log(O/H)+12 $\leq$ 7.35; \citealt{mcquinn2020leoncino}).  The galaxy found to have the lowest metallicity is AGC 123350 with a strong-line abundance of log(O/H)+12 = 7.46 $\pm$ 0.10.  This system has one of the lowest \ion{H}{1} masses found in our sample.  Overall, the galaxies in our ALFALFA-based sample possess abundances that are completely consistent with those measured in other samples of dwarf star-forming galaxies.  In particular, they overlap nicely with the LVL sample of \citet{berg2012} and the H$\alpha$ Dots from \citet{hirschauer2022} in the LZR plot shown in Figure~\ref{fig:LZR}.  The one exception is AGC 122939, which has characteristics that suggest that it may be of a tidal origin.

While our sample-selection method did not uncover any new XMP galaxies, it did allow us to explore a part of the ALFALFA parameter space that has seen very little attention: the subsample of low SNR \ion{H}{1} detections.  The fact that our small sample of measured metal abundances are completely consistent with those found from optically-selected samples is reassuring, and indicates that there are no significant biases in terms of metal content when selecting galaxies via their \ion{H}{1}.   Furthermore, it suggests that exploring the metallicities of ALFALFA galaxies with the lowest \ion{H}{1} masses may well lead to positive results in the search for additional XMP galaxies.  The ongoing spectroscopic study of the full SHIELD sample described above may well result in new examples of very metal-poor systems.

\vskip 0.32in
The operation of the WIYN 0.9 m telescope is supported by the College of Arts and Sciences at Indiana University.  The work presented here represents the Senior Honors Thesis research of JHM. The imaging portion of this study was carried out as a class project for the {\it Modern Astronomical Techniques} course offered by the Department of Astronomy at Indiana University.  Nearly a dozen undergraduate students were able to travel to Kitt Peak and gain valuable hands-on experience while helping to collect the observations used in this study.  This opportunity was made possible through the generous gift by William J. Delaney to support undergraduate research at IU.  We thank undergraduate students Matthew Britzman, Jacob Garcia, Elijah Guess, Jude Gussman, Garrett Johnson, Ruhee Kahar, Dustin Ruby, Madeline Shepley, Maria del Valle Coello, Nadine Williams for their participation in the project.  We also acknowledge graduate assistants Bobby Butler and Ryan Lambert, whose contributions made the class project possible.  MPH acknowledges support from grant NSF/AST-1714828 as well as from the Brinson Foundation.   Finally, we acknowledge several helpful suggestions made by the anonymous referee.  In particular, they suggested exploring the high N/O ratio explanation for the elevated abundances discussed in Section 6.1.

The Hobby-Eberly Telescope (HET) is a joint project of the University of Texas at Austin, the Pennsylvania State University, Ludwig-Maximilians-Universit\"at M\"unchen, and Georg-August-Universit\"at G\"ottingen.  The HET is named in honor of its principal benefactors, William P. Hobby and Robert E. Eberly.

This project made use of the Sloan Digital Sky Survey (SDSS).  Funding for the SDSS has been provided by the Alfred P. Sloan Foundation, the Participating Institutions, the National Science Foundation, the U.S. Department of Energy, the National Aeronautics and Space Administration, the Japanese Monbukagakusho, the Max Planck Society, and the Higher Education Funding Council for England. The SDSS Web Site is http://www.sdss.org/.  The SDSS is managed by the Astrophysical Research Consortium for the Participating Institutions. The Participating Institutions are the American Museum of Natural History, Astrophysical Institute Potsdam, University of Basel, University of Cambridge, Case Western Reserve University, University of Chicago, Drexel University, Fermilab, the Institute for Advanced Study, the Japan Participation Group, Johns Hopkins University, the Joint Institute for Nuclear Astrophysics, the Kavli Institute for Particle Astrophysics and Cosmology, the Korean Scientist Group, the Chinese Academy of Sciences (LAMOST), Los Alamos National Laboratory, the Max-Planck-Institute for Astronomy (MPIA), the Max-Planck-Institute for Astrophysics (MPA), New Mexico State University, Ohio State University, University of Pittsburgh, University of Portsmouth, Princeton University, the United States Naval Observatory, and the University of Washington.

%

\vspace{5mm}
\facilities{WIYN 0.9 m telescope, Hobby-Eberly Telescope}








\bibliographystyle{aasjournal}
\bibliography{bibliography.bib}

\end{document}